\documentclass[journal]{IEEEtran}
\usepackage{amsmath,amsfonts}
\usepackage{array}

\usepackage{textcomp}
\usepackage{stfloats}
\usepackage{url}
\usepackage{graphicx}
\usepackage{cite}
\usepackage[caption=false]{subfig}

\usepackage{booktabs}
\usepackage{multirow}
\usepackage{enumitem}
\usepackage{hyperref}

\usepackage{xcolor}

\definecolor{bestblue}{RGB}{0,153,204}
\newcommand{\best}[1]{\textcolor{bestblue}{\mathbf{#1}}}

\begin{document}

\title{Hadamard-Domain Model Quantization for Learned Image Coding}

\author{Junqi Shi, Chongzhi Wang, Yiwen He, Ming Lu, and Zhan Ma
\thanks{J. Shi, C. Wang, Y. He, M. Lu, and Z. Ma are with the School of Electronic Science and Engineering, Nanjing University, Nanjing, Jiangsu 210023, China (e-mail: junqishi@smail.nju.edu.cn; 522024230061@smail.nju.edu.cn; yiwenhe2022@smail.nju.edu.cn; minglu@nju.edu.cn; mazhan@nju.edu.cn).}
\thanks{J. Shi and C. Wang contributed equally to this work.}
}

\markboth{Journal of \LaTeX\ Class Files,~Vol.~14, No.~8, August~2021}%
{Shell \MakeLowercase{\textit{et al.}}: A Sample Article Using IEEEtran.cls for IEEE Journals}



\maketitle

\begin{abstract}

Uniform INT8 quantization is attractive for deploying learned image coding (LIC), but its rate--distortion (R--D) performance is often limited by heavy-tailed tensors and large inter-channel variations. Existing methods mainly adapt the quantizer through mixed precision or non-uniform codebooks. We propose \textit{Ha}damard-\textit{T}ransform-domain \textit{Q}uantization (\textit{HaTQ}), which uses orthogonal Hadamard reparameterization before quantization to redistribute weight and activation responses in the original domain across channels. The reparameterization preserves the original function mapping of each linear operator, while making its weights and activations more amenable to uniform INT8 quantization. HaTQ provides two complementary forms. Double-Hadamard quantization transforms both the input activations and weights, whereas weight-only Hadamard quantization transforms only the weights. This distinction is important because the constant Hadamard basis can coherently accumulate a nonzero channel mean and enlarge the activation range in sensitive layers. We identify these sensitive layers through offline profiling and assign the appropriate form to each layer without input-dependent branching. HaTQ supports both post-training quantization (PTQ) and quantization-aware training (QAT), uses uniform INT8 quantizers, and is compatible with integer-only execution. Experiments on representative LIC architectures and datasets demonstrate consistent improvements across different quantization settings. The resulting QAT models further outperform competing mixed-precision and non-uniform quantization methods. TensorRT deployment results demonstrate practical INT8 inference efficiency. The source code will be publicly released.
\end{abstract}


\section{Introduction}
\IEEEPARstart{L}{earned} image coding (LIC) replaces hand-crafted coding tools~\cite{wallace2002jpeg, sullivan2012overview, bross2021overview} with jointly optimized neural transforms and entropy models, enabling compressed representations tailored to reconstruction, perceptual quality, and downstream machine analysis~\cite{balle2018efficient,minnen2018joint,cheng2020learned,he2022elic,liu2023learned, shi2026dit, cong2026taming}. Recent standardization efforts, including JPEG AI~\cite{JPEG-AI} and IEEE Std 1857.11~\cite{IEEE1857.11}, further indicate its transition toward practical visual communication systems. However, the growing complexity of LIC models makes efficient and deterministic deployment as important as floating-point rate–-distortion (R--D) performance. 

Model quantization~\cite{shi2023rate, nagel2021white, yu2025activation} provides a natural solution by representing weights and intermediate activations with low-precision integers, thereby reducing storage, memory traffic, and arithmetic cost while limiting platform-dependent numerical variation. Such numerical consistency is particularly critical for LIC, where small encoder–decoder discrepancies in entropy-probability estimation may lead to arithmetic decoding failure~\cite{he2022post, koyuncu2022device, koyuncu2024quantized}. Therefore, practical LIC quantization must preserve end-to-end R--D performance while maintaining a regular low-precision computation path with reproducible deployment.

\begin{figure}[t]
    \centering
    \includegraphics[width=0.48\textwidth]{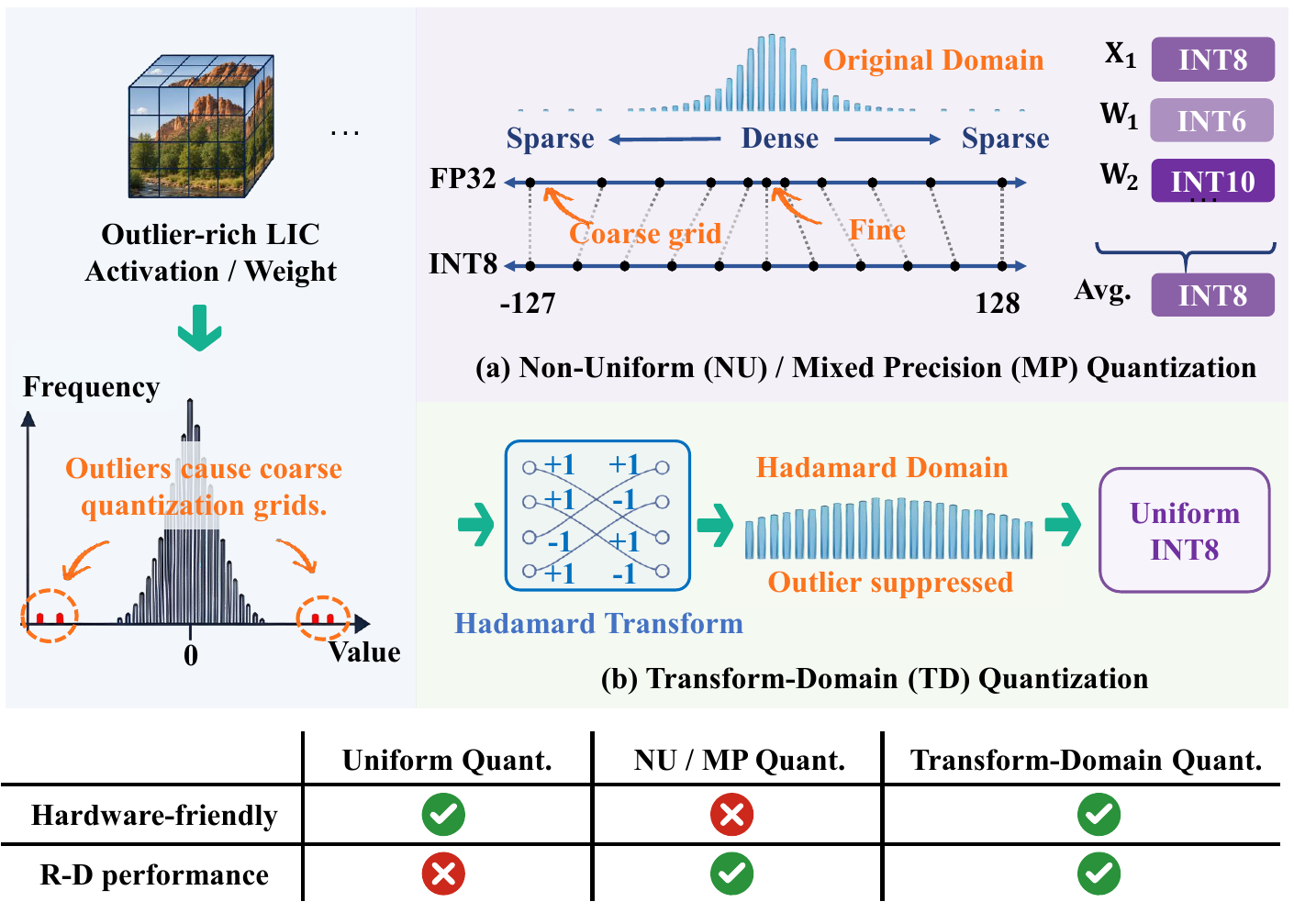}
    \caption{Overview of Hadamard transform-domain quantization for LIC. Outlier-rich weights and activations make original-domain uniform quantization hardware-friendly but vulnerable to severe R--D degradation, whereas non-uniform or mixed-precision quantization often improves performance with more complex inference. By equivalently transforming tensors into the Hadamard domain, our HaTQ suppresses outliers and enables uniform INT8 quantization to better balance R--D performance and deployment efficiency.}
    \label{fig:overview}
\end{figure}

Recent studies have explored QAT and PTQ for LIC~\cite{hong2020efficient, shi2023rate, sun2022q, sun2020end, sun2021fixed, yu2025activation, yang2025subset, zhang2026mpp, bao2026dynaquant}. Nevertheless, substantial R--D degradation remains under uniform INT8 quantization, particularly when activations are quantized tensor-wise for compatibility with common inference engines. One key difficulty lies in the numerical distributions of LIC tensors. Weights and activations often exhibit heavy-tailed responses and pronounced inter-channel magnitude variation, as illustrated in Fig.~\ref{fig:hadamard_visualization}. A few extreme values can dominate the quantization range and force most values to share a coarse quantization grid. This problem is especially pronounced in LIC, where latent representations are quantized into discrete symbols for entropy coding and the associated scaling and integer-domain representation further enlarge numerical ranges and magnitude variations.

Existing LIC model quantization methods address these distributional and sensitivity issues by adapting the quantizer or the precision assignment. Mixed-precision schemes~\cite{hossain2024flexible,yu2025mixed,zhang2026mpp, bao2026dynaquant} allocate additional bits to sensitive layers, whereas non-uniform schemes~\cite{yang2025subset, baskin2021uniq} place quantization grids unevenly across the value range to provide finer resolution for selected values. These methods improve accuracy, but the added flexibility may introduce heterogeneous bit widths, specialized numerical representations, or operators that are less directly supported by standard INT8 inference backends. This raises a complementary question: \textbf{\textit{instead of making the quantizer increasingly adaptive, can we make the tensors themselves easier to quantize while keeping the quantizer simple?}}

We answer this question by revisiting a fundamental compression principle: transform coding~\cite{clarke1985transform,goyal2001theoretical}, which maps signals into domains that permit lower-distortion quantization and more efficient entropy coding. While traditionally applied to source signals such as images and videos, we extend this principle to model weights and activations by seeking function-preserving reparameterizations that better align their distributions with a uniform integer grid. 
The Hadamard transform is particularly attractive for this purpose because it is orthogonal, sign-structured, inexpensive to implement, and has been used in transform coding since the late 1960s~\cite{pratt1969hadamard}. Each transformed channel is formed from an equal-magnitude signed combination of the original channels, thereby redistributing channel-localized responses across channels. Moreover, when Hadamard transforms are introduced in algebraically paired forms, their factors cancel algebraically, preserving the input--output mapping of the original linear operator.

Building on this principle, we propose \textbf{\textit{Ha}}damard-\textbf{\textit{T}}ransform-domain \textbf{\textit{Q}}uantization (\textbf{\textit{HaTQ}}), a framework for LIC models. 
HaTQ quantizes Hadamard-domain weights and activations through two complementary operator
forms. \emph{Double-Hadamard (DH) quantization} applies matched transforms to the activation and weight operands along their contracted dimension. \emph{Weight-only Hadamard (WH) quantization} instead transforms the weights along the output-channel dimension and restores the original output basis after accumulation, thereby avoiding activation-side transformation. 

This distinction is necessary because Hadamard transformation is not uniformly beneficial for all layers. In most layers, it suppresses heavy-tailed responses and improves utilization of the INT8 grid. In a small set of activation-sensitive layers, however, nonlinearities can produce a nonzero mean across channels. The constant-basis component then coherently accumulates this mean, potentially creating an extreme response that expands the tensor-wise quantization range. We therefore develop an offline profiling strategy that selects DH or WH for each layer according to its activation-range expansion. The resulting assignment remains fixed during optimization and inference, introducing no input-dependent branching.

At its core, HaTQ applies the principle of transform coding to model quantization: a simple uniform quantizer can be more effective when weights and activations are in a more quantization-friendly representation domain. Our main contributions are summarized as follows:

\begin{enumerate}[label=\arabic*)]
\item \textbf{Hadamard-domain quantization framework.}
We introduce HaTQ, a transform-domain framework that incorporates function-preserving Hadamard reparameterizations into LIC linear operators. It supports uniform INT8 quantization under both PTQ and QAT, while remaining compatible with integer-only inference.

\item \textbf{Layer-adaptive Hadamard quantization.}
We characterize a coherent accumulation effect in which a nonzero activation mean produces a dominant constant-basis component and enlarges the tensor-wise quantization range. Based on this observation, we develop an offline profiling mechanism that assigns either DH or WH quantization to each layer.

\item \textbf{Comprehensive evaluation and deployment.}
Experiments across representative LIC architectures, datasets, and quantization granularities demonstrate consistent R--D improvements over existing quantization methods. A TensorRT implementation further evaluates the practical inference latency and storage overhead of HaTQ under INT8 deployment.
\end{enumerate}

\section{Related Work}

\subsection{Learned Image Coding}
High-dimensional vector quantization (VQ) can, in principle, approach the rate--distortion (R--D) limit of a source, but its codebook size and search complexity grow prohibitively with the source dimension~\cite{gray1984vector,gersho2012vector,balle2020nonlinear}. Modern image codecs therefore predominantly follow the transform-coding paradigm, which maps images into compact representations for tractable quantization and entropy coding~\cite{goyal2001theoretical,jayant1984digital}. Classical systems instantiate this principle with analytically designed tools, such as Huffman coding~\cite{huffman1952method} and the discrete cosine transform~\cite{DCT}. LIC inherits this transform-coding view but replaces fixed transforms, quantizers, and probability models with learned data-driven modules.

Early neural image-coding studies explored the use of neural networks for compact representations~\cite{daugman1988complete,cottrell1987image,dony1995neural,jiang1999image}, but the modern LIC paradigm was established when nonlinear transforms and entropy models became trainable end to end. Toderici et al.~\cite{toderici2015variable} demonstrated variable-rate recurrent coding, while Ball{\'e} et al.~\cite{balle2017end,balle2018efficient} connected nonlinear analysis/synthesis transforms, differentiable quantization surrogates, and learned entropy models into a principled R--D optimization framework. Minnen et al.~\cite{minnen2018joint, minnen2020channel} further showed that hierarchical priors and autoregressive contexts can substantially improve entropy estimation. These works shifted learned coding from isolated neural components to an end-to-end nonlinear transform-coding system.

Later work has expanded this framework along two main axes. The first improves the analysis and synthesis transforms through residual, attention-based, transformer, and state-space architectures~\cite{cheng2019deep,zou2022devil,liu2023learned,feng2025linear,zeng2025mambaic}. The second improves probability modeling through hyperpriors, spatial/channel contexts, and transformer-based entropy models~\cite{balle2018efficient,li2018learning,Qian2022Entroformer,koyuncu2022contextformer}. Hierarchical and conditional formulations further enable progressive reconstruction and flexible rate control~\cite{nakanishi2018neural,duan2023qarv,zhang2026qarv++, cong2026taming}, while learned representations also support perceptual, machine-oriented, and joint source--channel objectives~\cite{dubois2021lossy,bourtsoulatze2019deep, chen2023transtic, zhang2025perception}. More recently, generative priors, including diffusion models, have been incorporated as expressive decoders to improve perceptual reconstruction at very low bitrates~\cite{yang2023lossy, relic2024lossy, shi2026dit, li2026yoda, li2024toward}. As visual data increasingly serves both human perception and machine intelligence, this flexibility positions LIC as a key enabling technology for AI-era visual communication. At the same time, increasingly sophisticated transforms, entropy models, and iterative generative decoders further increase computation, memory traffic, and inference latency. Deployment-oriented optimization, including low-precision execution, is therefore becoming an increasingly central issue for LIC.

\subsection{Model Quantization}
Model quantization reduces storage, memory traffic, and arithmetic cost by representing weights and activations with low-precision numbers. 
It should be distinguished from latent quantization in learned codecs, which produces discrete symbols for entropy coding and directly controls the coding rate. The efficiency potential is substantial: under a 45-nm CMOS estimate, an 8-bit integer multiplication consumes about 0.2~pJ, compared with 3.7~pJ for a 32-bit floating-point multiplication~\cite{horowitz2014computing}. Although the actual savings depend on the hardware architecture and fabrication technology, this comparison illustrates the efficiency potential of low-precision computation.

Quantization methods are often distinguished by their optimization regime. Post-training quantization (PTQ) converts a pretrained model using a small calibration set, with quantization parameters commonly determined from numerical ranges or by minimizing quantization or task error~\cite{choukroun2019low,bhandare2019efficient, nagel2021white}. It is deployment-friendly but has limited ability to absorb quantization error. Quantization-aware training (QAT) simulates low-precision arithmetic during training, allowing weights and quantization parameters to adapt to the target precision at the cost of additional training~\cite{wei2024advances, nagel2021white}.

Orthogonal to this PTQ/QAT distinction is the quantizer design. Uniform INT8 quantization is attractive because it maps naturally to widely available integer kernels, but it is vulnerable to outliers and inter-channel range imbalance. Mixed-precision methods allocate more bits to sensitive layers or channels~\cite{dettmers2022gpt3, shi2025quantizing, dong2019hawq}, whereas non-uniform methods adapt quantization levels or numerical formats to the underlying distributions~\cite{noune20228, baskin2021uniq, kim2023squeezellm}.

Another line of work changes the numerical domain before quantization without changing the full-precision function. SmoothQuant~\cite{xiao2023smoothquant} uses channel-wise scaling to migrate activation outlier difficulty to weights, making activations easier to quantize with uniform grids. Rotation-based approaches go further by redistributing numerical energy across dimensions through orthogonal transformations: QuaRot~\cite{ashkboos2024quarot} applies randomized Hadamard rotations, SpinQuant~\cite{liu2024spinquant} optimizes learnable rotations, and QuIP\#~\cite{tseng2024quip} introduces lattice codebooks for weight quantization. These methods reveal an important principle: quantization quality is determined not only by the quantizer, but also by the coordinate system in which tensors are quantized.

\subsection{LIC Model Quantization}
\label{sec:model_quant}


Early studies primarily focused on QAT. Ball{\'e} et al.~\cite{balle2018integer} redesigned learned codecs for integer arithmetic, and subsequent work extended low-precision computation from weights to activations and nonlinear modules~\cite{sun2020end,sun2021fixed}. Other methods then improved R--D robustness by exposing the codec to quantization effects during optimization. RAQ~\cite{hong2020efficient}, Q-LIC~\cite{sun2022q}, and IQ-LIC~\cite{jeon2023integer} introduced range adaptation, sensitivity-aware channel redistribution, and quantization-aware objectives, respectively. Quantization has also been combined with pruning and variable-rate coding~\cite{hossain2024structured, ye2025variable}. Although QAT can substantially recover R--D performance, it requires access to training data and costly codec-level retraining.

PTQ offers a lighter path for converting pretrained LIC models. Cross-platform PTQ and quantized-decoder studies focus on entropy-model consistency and deterministic reconstruction, showing that carefully quantizing entropy-related modules can preserve codec interoperability~\cite{he2022post,koyuncu2022device,koyuncu2024quantized,JPEG-AI}. RDO-PTQ further demonstrates that calibration should optimize the end-to-end R--D objective rather than local tensor reconstruction alone~\cite{shi2023rate}.

\begin{figure*}[t]
\centering
\subfloat[Original Activation Domain \label{fig:hadamard_activation_original}]{%
\includegraphics[width=0.24\textwidth]{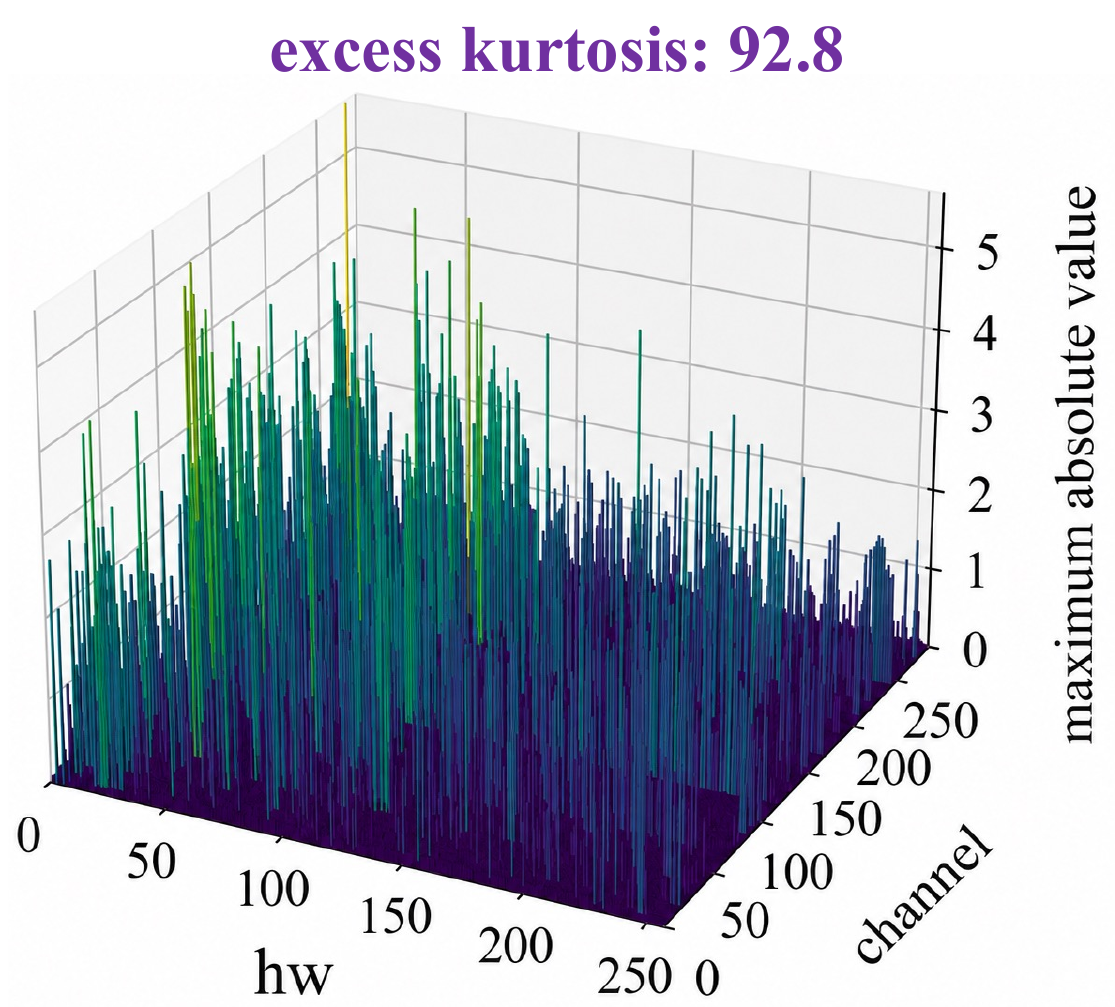}%
}
\subfloat[Hadamard Activation Domain \label{fig:hadamard_activation_transformed}]{%
\includegraphics[width=0.24\textwidth]{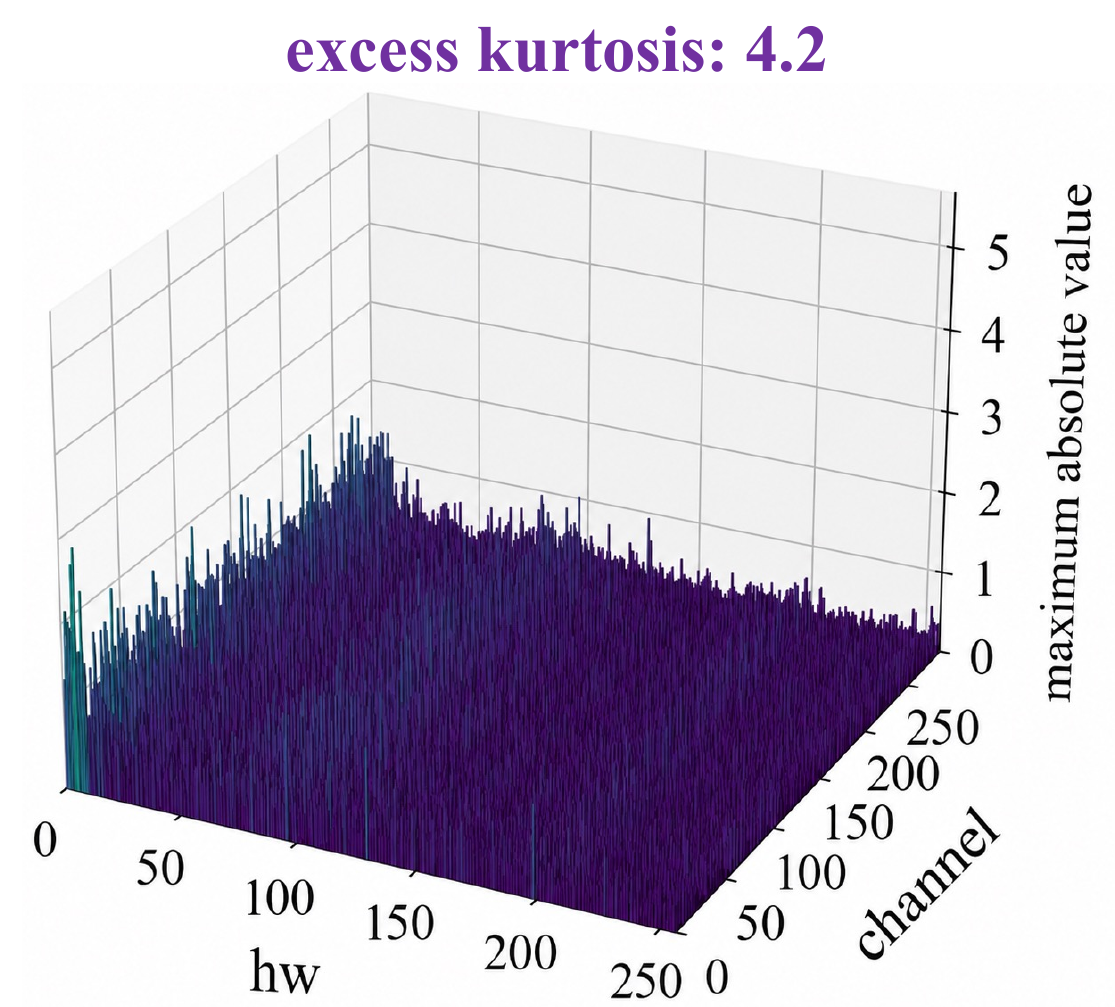}%
}
\subfloat[Original Weight Domain \label{fig:hadamard_weight_original}]{%
\includegraphics[width=0.24\textwidth]{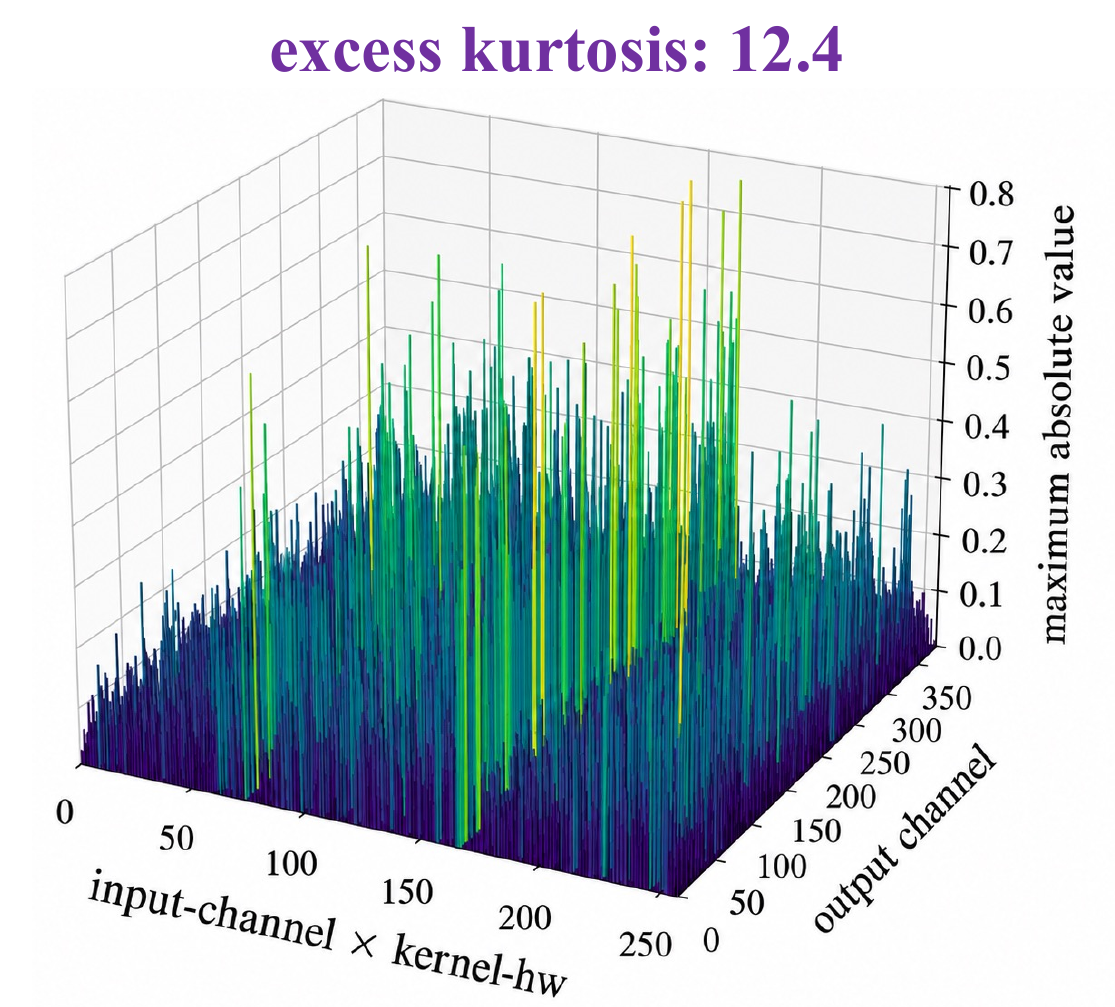}%
}
\subfloat[Hadamard Weight Domain \label{fig:hadamard_weight_transformed}]{%
\includegraphics[width=0.24\textwidth]{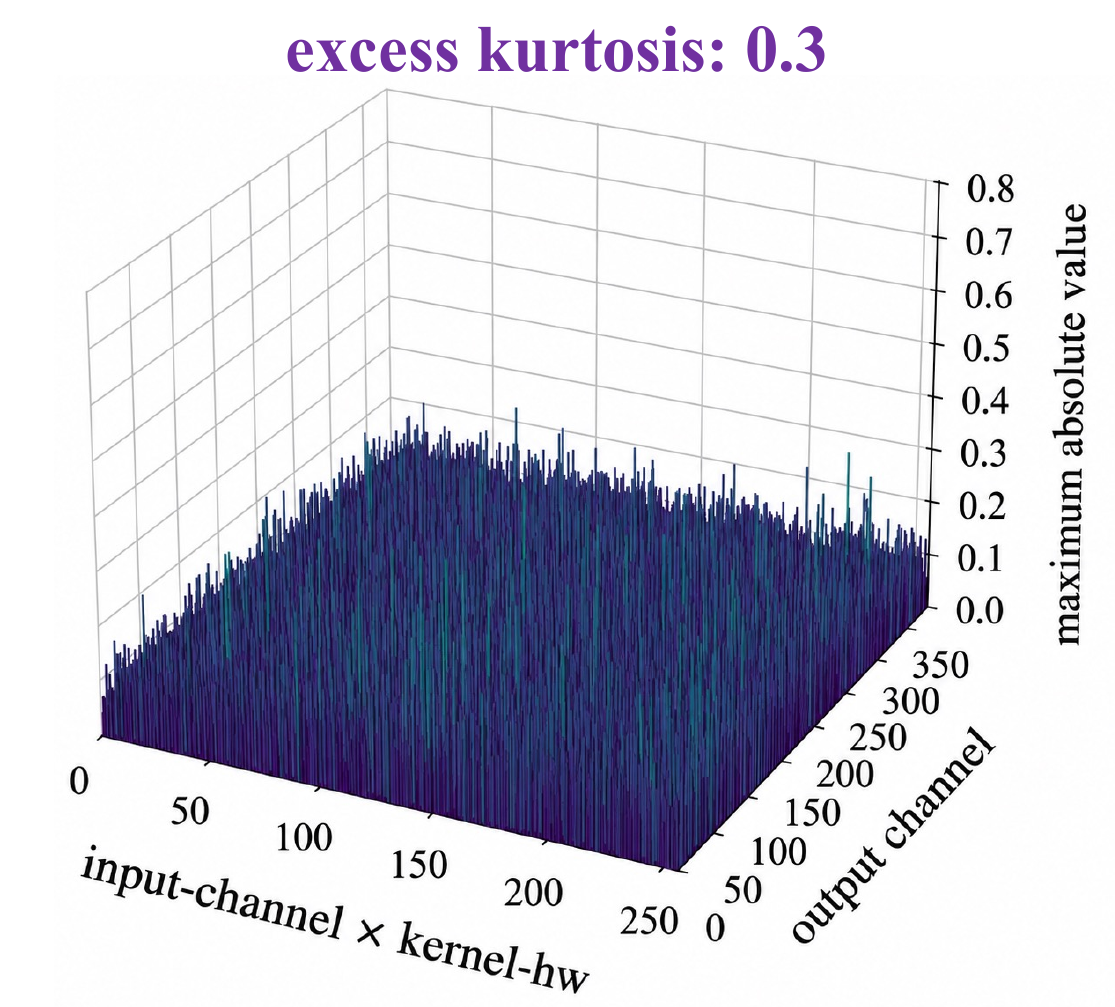}%
}
\caption{Absolute magnitudes of the input activations and weights in the original and Hadamard domains. The transform substantially suppresses outliers, reducing the excess kurtosis of activations and weights by 95.5\% and 97.6\%, respectively.}
\label{fig:hadamard_visualization}
\end{figure*}

More recent methods explicitly exploit the heterogeneous quantization sensitivity of LIC networks. FMPQ~\cite{hossain2024flexible} assigns layer-wise bit-widths according to the fractional increase in R--D loss and employs an adaptive search to meet a given model-size budget, whereas MP-PTQ~\cite{yu2025mixed} evaluates layer sensitivity through end-to-end R--D analysis and jointly determines the bit-width and quantization scale from a small calibration set. DynaQuant~\cite{bao2026dynaquant} further extends static layer-wise allocation with content-adaptive scales and offsets, together with an input-dependent selector that dynamically adjusts the precision of individual layers. AWDB~\cite{yu2025activation} addresses outliers and range imbalance by transferring quantization difficulty between paired activations and weights or redistributing salient values across channels, thereby reducing clipping and rounding errors under low-precision quantization. SS-PTQ selects distribution-matched quantization levels from a dense candidate pool, providing a non-uniform quantizer that better represents irregular layer- or channel-wise weight distributions~\cite{yang2025subset}.

These quantizer-oriented designs improve quantization robustness, but their numerical flexibility does not necessarily translate into efficient end-to-end inference. Heterogeneous bit widths, non-uniform representations, and structural modifications may require precision conversions, specialized kernels, or irregular execution paths. Such requirements complicate operator fusion and scheduling, fragment the codec pipeline, and limit direct use of the uniform INT8 kernels available on commodity accelerators. In contrast, our transform-domain approach redistributes tensors with equivalent Hadamard transforms before quantization, retaining a regular uniform INT8 computation path without heterogeneous precision or non-uniform arithmetic. This design targets both R--D robustness and compatibility with existing low-precision hardware and deployment pipelines.

\section{Preliminaries}
\subsection{Uniform Quantization}

Uniform quantization~\cite{nagel2021white} maps a real-valued tensor $x$ to an integer grid. Given a scale $s>0$, a zero-point $z$, and an integer range $[q_{\min},q_{\max}]$, the element-wise quantization and dequantization operations are
\begin{align}
x_{\mathrm{int}}
&=
\operatorname{clamp}
\left(
\operatorname{round}\left(\frac{x}{s}\right)+z,\,
q_{\min},q_{\max}
\right), \nonumber\\
\hat{x}
&=
s(x_{\mathrm{int}}-z).
\label{eq:uniform_q}
\end{align}
For affine min--max quantization, let
\(\bar{x}_{\min}=\min(x_{\min},0)\) and
\(\bar{x}_{\max}=\max(x_{\max},0)\), so that zero is representable. For a nondegenerate range, the quantization parameters are
\begin{align}
s&=\frac{\bar{x}_{\max}-\bar{x}_{\min}}{q_{\max}-q_{\min}}, \nonumber\\
z&=\operatorname{clamp}
\left(
\operatorname{round}\left(q_{\min}-\frac{\bar{x}_{\min}}{s}\right),
q_{\min},q_{\max}
\right).
\label{eq:affine_q}
\end{align}
For unsigned $b$-bit quantization, $q_{\min}=0$ and $q_{\max}=2^b-1$.

Static quantization fixes $(s,z)$ after calibration or training, whereas dynamic quantization computes them from each runtime activation tensor. In our PTQ setting, weights use static quantization and activations use dynamic affine quantization. For QAT, we use static symmetric quantization. Defining \(q_{\max}^{\mathrm{s}}=2^{b-1}-1\), we have
\begin{align}
x_{\mathrm{int}}
&=
\operatorname{clamp}
\left(
\operatorname{round}\left(\frac{x}{s}\right),
-q_{\max}^{\mathrm{s}},\,q_{\max}^{\mathrm{s}}
\right), \nonumber\\
s&=\frac{\|x\|_{\infty}}{q_{\max}^{\mathrm{s}}},
\qquad z=0.
\label{eq:symmetric_q}
\end{align}

Symmetric quantization removes zero-point corrections and simplifies integer accumulation. However, extreme values increase $s$ and coarsen the quantization grid available to the remaining values. This sensitivity motivates the distribution reshaping developed in Sec.~\ref{sec:hadamard_quantization}.

\subsection{Integer-Only Inference}
\label{sec:integer_only_inference}

Eq.~\eqref{eq:uniform_q} gives the dequantized values used to simulate quantization during PTQ or QAT. At inference, a linear or convolutional layer can instead accumulate directly on the integer codes. Suppressing the output indices and summing over the contracted dimension \(i\), the accumulator is
\begin{align}
a_{\mathrm{int}}
&=
\sum_i
(x_{\mathrm{int},i}-z_x)
(w_{\mathrm{int},i}-z_w)
+b_{\mathrm{int}}, \nonumber\\
b_{\mathrm{int}}
&=
\operatorname{round}\left(\frac{b}{s_xs_w}\right).
\label{eq:integer_accumulation}
\end{align}
Since $a_{\mathrm{int}}$ has scale $s_xs_w$, conversion to an output grid with scale $s_y$ requires the ratio $s_xs_w/s_y$. This ratio is approximated by an integer multiplier $M$ and a nonnegative shift $n$:
\begin{equation}
\label{eq:fixed_point_scale}
\frac{s_xs_w}{s_y}
\approx
\frac{M}{2^n}.
\end{equation}
The output code is then obtained by requantization:
\begin{equation}
\label{eq:integer_requantization}
y_{\mathrm{int}}
=
\operatorname{clamp}
\left(
\operatorname{round}
\left(
\frac{M a_{\mathrm{int}}}{2^n}
\right)
+z_y,\,
q_{\min},q_{\max}
\right).
\end{equation}
Thus, the layer uses integer multiply--accumulate, fixed-point rescaling, rounding, and saturation; no floating-point tensor need be materialized between compatible quantized operators. In Sec.~\ref{sec:hadamard_quantization}, we show that the proposed Hadamard-domain operators preserve this integer-only execution form.

\subsection{Hadamard Transform}

Let $\widetilde{H}\in\{-1,+1\}^{m\times m}$ be a Hadamard matrix~\cite{pratt1969hadamard, seberry2020hadamard} satisfying $\widetilde{H}\widetilde{H}^{\top}=mI$. Its normalized form is
\begin{equation}
\label{eq:hadamard_properties}
H=\frac{1}{\sqrt{m}}\widetilde{H},
\qquad HH^{\top}=H^{\top}H=I.
\end{equation}
Hence, $H^{-1}=H^{\top}$ and the transform preserves inner products and Euclidean norms. For a vector $x\in\mathbb{R}^{m}$, each component of $y=Hx$ is a normalized signed sum:
\begin{equation}
\label{eq:hadamard_mixing}
y_i=\frac{1}{\sqrt{m}}\sum_{j=1}^{m}\widetilde{H}_{ij}x_j.
\end{equation}
Each input component therefore contributes equally in magnitude to every transformed component. It can disperse channel-localized extremes and yield a range better matched to uniform quantization. Sec.~\ref{sec:hadamard_quantization} uses the orthogonality of $H$ to transform weights and activations without changing the function mapping of the original operator.

\section{Methodology}

\begin{figure}[ht]
    \centering
    \includegraphics[width=0.85\columnwidth]{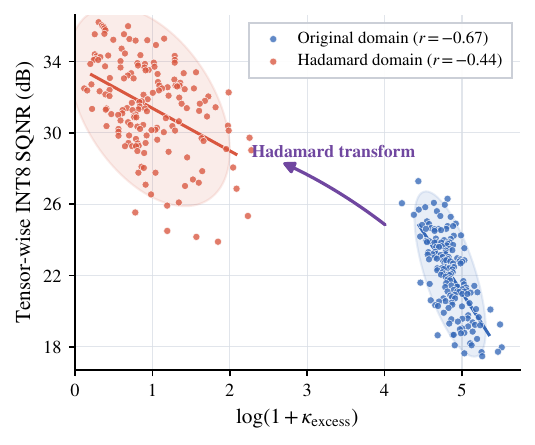}
    \caption{Activation excess kurtosis versus SQNR under tensor-wise INT8 quantization. Each point denotes the input-activation tensor produced by one test image, evaluated either in the original domain or after a Hadamard transform. Solid lines are least-squares fits, and shaded ellipses indicate 90\% covariance regions. The Hadamard-domain samples generally exhibit lower excess kurtosis and higher SQNR.}
    \label{fig:sqnr}
\end{figure}

\subsection{From Original-Domain to Transform-Domain Quantization}
\label{sec:transform_domain_motivation}

\textbf{Outlier-dominated distributions in the original domain.}
LIC weights and activations commonly contain a dense concentration of small values and a sparse set of large responses, as illustrated in Fig.~\ref{fig:hadamard_visualization}(a) and (c). Under uniform quantization, these responses determine the quantization range and hence the step size, leaving fewer effective grids for the bulk of the tensor. This limitation is particularly severe for per-tensor quantization because one range is shared across all channels and spatial positions. Channel-wise quantization mitigates inter-channel range imbalance, but remains sensitive to extreme values within each channel.

This distributional mismatch leads to substantial R--D degradation. For example, the original-domain QAT baseline with tensor-wise activations and channel-wise weights incurs an 18.63\% BD-rate increase; the complete ablation setting is reported in Fig.~\ref{fig:hadamard_ablation}. This result indicates that the original-domain distributions of LIC tensors are not well suited to low-precision uniform quantization.

We quantify tail heaviness by the empirical \textit{excess kurtosis}~\cite{decarlo1997meaning, hatem2022normality}. For a nonconstant tensor \(X=\{x_i\}_{i=1}^{N}\),
\begin{align}
\mu_X
&=
\frac{1}{N}\sum_{i=1}^{N}x_i,
\nonumber\\
\kappa_{\mathrm{ex}}(X)
&=
\frac{
\frac{1}{N}\sum_{i=1}^{N}(x_i-\mu_X)^4
}{
\left[
\frac{1}{N}\sum_{i=1}^{N}(x_i-\mu_X)^2
\right]^2
}
-3.
\label{eq:excess_kurtosis}
\end{align}
The subtraction of \(3\) sets the Gaussian reference value to zero: a Gaussian distribution has excess kurtosis \(0\). A larger \(\kappa_{\mathrm{ex}}(X)\) indicates greater fourth-moment concentration and, typically, stronger sensitivity to isolated large-magnitude values. Quantization fidelity is measured using the signal-to-quantization-noise ratio (SQNR):
\begin{equation}
\operatorname{SQNR}(X,\widehat{X})
=
10\log_{10}
\frac{\lVert X\rVert_F^2}
{\lVert X-\widehat{X}\rVert_F^2},
\label{eq:sqnr}
\end{equation}
where \(\widehat{X}\) is the dequantized tensor. A higher SQNR indicates a smaller relative quantization error. Fig.~\ref{fig:sqnr} shows a negative correlation between activation excess kurtosis and SQNR, linking heavy-tailed activation distributions to poor utilization of the uniform integer grid.

\textbf{Distribution reshaping in the Hadamard domain.}
We therefore apply an orthogonal Hadamard transform before quantization. Its equal-magnitude signed mixing disperses responses localized to individual channels while preserving the tensor Frobenius norm. In the representative layer of Fig.~\ref{fig:hadamard_visualization}, this transformation produces more balanced activation and weight magnitudes and substantially reduces their excess kurtosis. Across various test images, Fig.~\ref{fig:sqnr} further shows that the transformed activations generally attain higher SQNR.

These results motivate quantizing LIC in the Hadamard domain rather than changing the uniform quantizer itself. Orthogonality alone preserves tensor energy, but network operator equivalence requires the activation and weight transforms to be paired appropriately. The next subsection derives these equivalent constructions.

\subsection{Hadamard Transform-Domain LIC Quantization}
\label{sec:hadamard_quantization}

\begin{figure}[t]
    \centering
    \includegraphics[width=0.48\textwidth]{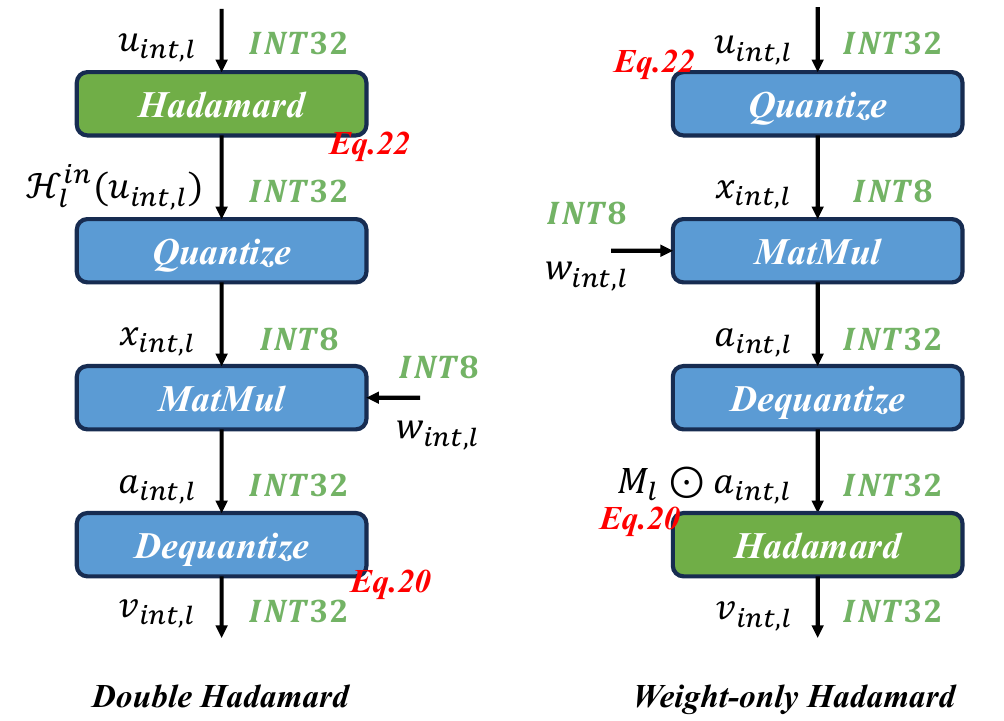}
    \caption{Integer inference paths of double-Hadamard (DH, left) and weight-only Hadamard (WH, right) quantization. DH transforms the activation into the input Hadamard basis before quantization and INT8 matrix multiplication. WH keeps the activation in its original basis and restores the output basis after accumulation. The blocks labeled \textit{Quantize} and \textit{Dequantize} perform fixed-point integer rescaling, and they do not materialize floating-point tensors.}
    \label{fig:hadamard_framework}
\end{figure}

We construct two equivalent reparameterizations for transform-domain quantization. Double-Hadamard (DH) quantization transforms both weights and activations along their contracted dimension, whereas weight-only Hadamard (WH) quantization eliminates the activation-side transform by applying a Hadamard transform to the weight output channels and subsequently restoring the original output basis. Fig.~\ref{fig:hadamard_framework} summarizes their integer inference paths.

\textbf{Hadamard-domain formulation of linear operators.}
After unfolding a convolution into input patches, both convolutional and fully connected layers can be written as
\begin{equation}
Y_l=X_lW_l^{\top},
\label{eq:linear_operator}
\end{equation}
where \(X_l\in\mathbb{R}^{N_l\times D_l}\) contains the input vectors, \(W_l\in\mathbb{R}^{C_{\mathrm{out},l}\times D_l}\) contains the weight vectors, and \(Y_l\in\mathbb{R}^{N_l\times C_{\mathrm{out},l}}\). Here, \(D_l=C_{\mathrm{in},l}K_hK_w\) for a \(K_h\times K_w\) convolution and \(D_l=C_{\mathrm{in},l}\) for a fully connected layer. Bias terms are omitted for readability and are treated in supplementary materials.

Let \(\mathbf H_n\in\mathbb{R}^{n\times n}\) be a normalized Hadamard matrix satisfying \(\mathbf H_n\mathbf H_n^{\top}=I_n\). For clarity, we first assume that the channel dimensions admit square Hadamard matrices of the same orders; the padded construction is introduced in Sec.~\ref{sec:hadamard_construction}. The input-side transform is
\begin{equation}
\mathbf H_l^{\mathrm{in}}
=
\begin{cases}
I_{K_hK_w}\otimes\mathbf H_{C_{\mathrm{in},l}}, & \text{convolution},\\
\mathbf H_{C_{\mathrm{in},l}}, & \text{fully connected},
\end{cases}
\label{eq:input_hadamard_transform}
\end{equation}
where \(\otimes\) denotes the Kronecker product. For convolution, this block-diagonal form applies the same channel transform independently at each kernel position and need not be explicitly constructed.

\textbf{Double-Hadamard quantization.}
Since \(\mathbf H_l^{\mathrm{in}}(\mathbf H_l^{\mathrm{in}})^{\top}=I_{D_l}\), Eq.~\eqref{eq:linear_operator} is equivalently expressed as
\begin{equation}
Y_l
=
\left(X_l\mathbf H_l^{\mathrm{in}}\right)
\left(W_l\mathbf H_l^{\mathrm{in}}\right)^{\top}.
\label{eq:hadamard_linear_equivalence}
\end{equation}
Quantizing both transformed operands gives
\begin{equation}
\widehat{Y}_l^{\mathrm{DH}}
=
Q_a\!\left(X_l\mathbf H_l^{\mathrm{in}}\right)
Q_w\!\left(W_l\mathbf H_l^{\mathrm{in}}\right)^{\top},
\label{eq:double_hadamard_quantization}
\end{equation}
where \(Q_a(\cdot)\) and \(Q_w(\cdot)\) denote quantization followed by dequantization for activations and weights, respectively. We call Eq.~\eqref{eq:double_hadamard_quantization} double-Hadamard (DH) quantization because both operands are quantized in the same input Hadamard basis. In the underlying full-precision identity, the paired transforms cancel through \(\mathbf H_l^{\mathrm{in}}(\mathbf H_l^{\mathrm{in}})^{\top}=I_{D_l}\); thus, no output-side basis conversion is required.

\textbf{Weight-only Hadamard quantization.}
We instead retain \(X_l\) in its original basis and transform \(W_l\) along the output-channel dimension. Define
\begin{equation}
\mathbf H_l^{\mathrm{out}}
=
\mathbf H_{C_{\mathrm{out},l}}
\in
\mathbb{R}^{C_{\mathrm{out},l}\times C_{\mathrm{out},l}}
\label{eq:output_hadamard_transform}
\end{equation}
as the normalized output-side transform. Orthogonality gives
\begin{equation}
Y_l
=
\left[
X_l
\left(
\mathbf H_l^{\mathrm{out}}W_l
\right)^{\top}
\right]
\mathbf H_l^{\mathrm{out}}
\label{eq:weight_only_hadamard_equivalence}
\end{equation}
and its quantized counterpart is
\begin{equation}
\widehat{Y}_l^{\mathrm{WH}}
=
\left[
Q_a(X_l)
Q_w\!\left(
\mathbf H_l^{\mathrm{out}}W_l
\right)^{\top}
\right]
\mathbf H_l^{\mathrm{out}}.
\label{eq:weight_only_hadamard_quantization}
\end{equation}
The matrix multiplication produces an accumulator in the output Hadamard basis, which is subsequently mapped back to the original basis by \(\mathbf H_l^{\mathrm{out}}\). Accordingly, \emph{weight-only} means that only the weights are transformed.

\textbf{Integer-only inference.}
Although Eq.~\eqref{eq:double_hadamard_quantization} and~\eqref{eq:weight_only_hadamard_quantization} are expressed in dequantized form, both operators support integer-only execution under the hardware-prevalent setting of symmetric tensor-wise activation quantization and channel-wise weight quantization.  Let \(s_{x,l}\) be the activation scale of the INT8 operand consumed by layer \(l\), and let \(s_{w,l,c}\) be the weight scale of output channel \(c\). The INT8 multiply--accumulate produces an INT32 tensor \(a_{\mathrm{int},l}\). Its \(c\)-th output channel has the accumulator scale
\begin{equation}
s_{A,l,c}=s_{x,l}s_{w,l,c}.
\label{eq:accumulator_scale}
\end{equation}
Thus, the real-valued accumulated output in channel \(c\) is approximated by \(s_{A,l,c}a_{\mathrm{int},l,c}\).

We denote the corresponding unnormalized integer output- and input-side transforms by \(\mathcal{H}_l^{\mathrm{out}}\) and \(\mathcal{H}_{l+1}^{\mathrm{in}}\), respectively. They require only additions and subtractions. Let \(m_l^{\mathrm{out}}\) and \(m_{l+1}^{\mathrm{in}}\) be the actual Hadamard orders, equal to the corresponding channel dimensions when no padding is used. Table~\ref{tab:integer_transitions} lists the required transforms and their normalization factors.

\begin{table}[t]
\centering
\caption{Integer basis conversions between adjacent DH and WH layers. The output-side transform \(\mathcal{H}_l^{\mathrm{out}}\) is applied before the activation function \(\phi\), and the input-side transform \(\mathcal{H}_{l+1}^{\mathrm{in}}\) is applied afterward. Here, \(\mathcal{I}\) is the identity and \(m\) denotes the actual Hadamard order, including any padded dimensions.}
\label{tab:integer_transitions}
\setlength{\tabcolsep}{4pt}
\renewcommand{\arraystretch}{1.10}
\begin{tabular}{cccc}
\toprule
\textbf{Transition}
&
\(\left(\gamma_l^{\mathrm{out}},\gamma_{l+1}^{\mathrm{in}}\right)\)
&
\shortstack{\(\mathcal{H}_l^{\mathrm{out}}\)\\before \(\phi\)}
&
\shortstack{\(\mathcal{H}_{l+1}^{\mathrm{in}}\)\\after \(\phi\)}
\\
\midrule
DH-to-WH
&
\((1,1)\)
&
\(\mathcal{I}\)
&
\(\mathcal{I}\)
\\
DH-to-DH
&
\(\left(1,\sqrt{m_{l+1}^{\mathrm{in}}}\right)\)
&
\(\mathcal{I}\)
&
\(\mathcal{H}_{l+1}^{\mathrm{in}}\)
\\
WH-to-WH
&
\(\left(\sqrt{m_l^{\mathrm{out}}},1\right)\)
&
\(\mathcal{H}_l^{\mathrm{out}}\)
&
\(\mathcal{I}\)
\\
WH-to-DH
&
\(\left(\sqrt{m_l^{\mathrm{out}}},\sqrt{m_{l+1}^{\mathrm{in}}}\right)\)
&
\(\mathcal{H}_l^{\mathrm{out}}\)
&
\(\mathcal{H}_{l+1}^{\mathrm{in}}\)
\\
\bottomrule
\end{tabular}
\end{table}

The factors \(\gamma_l^{\mathrm{out}}\) and \(\gamma_{l+1}^{\mathrm{in}}\) are the denominators of the normalized output- and input-side transforms in Table~\ref{tab:integer_transitions}. Their product and the scale ratio for accumulator channel \(c\) are
\begin{align}
\alpha_l
&=
\gamma_l^{\mathrm{out}}
\gamma_{l+1}^{\mathrm{in}},
\nonumber\\
\rho_{l,c}
&=
\frac{s_{A,l,c}}
{\alpha_l s_{x,l+1}}
\approx
\frac{M_{l,c}}{2^{n_l}},
\qquad
M_{l,c}
=
\operatorname{round}
\left(
2^{n_l}\rho_{l,c}
\right),
\label{eq:integer_fixed_point_scale}
\end{align}
where \(M_{l,c}\) is an integer multiplier and \(n_l\geq0\) is a shared right shift. Both are determined offline subject to multiplier precision and overflow constraints. The Hadamard normalization is therefore absorbed into fixed-point requantization.

For the identity mapping, or for a positively homogeneous activation such as ReLU, LeakyReLU, or PReLU, satisfying
\(\phi(\beta z)=\beta\phi(z)\) for \(\beta>0\), the common factor \(2^{-n_l}\) can be propagated through \(\phi\). The transition then becomes
\begin{align}
v_{\mathrm{int},l}
&=
\mathcal{H}_l^{\mathrm{out}}
\left(
\boldsymbol{M}_l\odot a_{\mathrm{int},l}
\right),
\\
u_{\mathrm{int},l}
&=
\phi_{\mathrm{int}}
\left(
v_{\mathrm{int},l}
\right),
\\
x_{\mathrm{int},l+1}
&=
\operatorname{clamp}
\left(
\operatorname{round}
\left[
\frac{
\mathcal{H}_{l+1}^{\mathrm{in}}
\left(
u_{\mathrm{int},l}
\right)
}
{2^{n_l}}
\right]
,
q_{\min},q_{\max}
\right),
\label{eq:integer_hadamard_transition}
\end{align}
where \(\phi_{\mathrm{int}}\) preserves the common implicit scale, \(\odot\) denotes channel-wise multiplication, and \(x_{\mathrm{int},l+1}\) is the INT8 operand consumed by layer \(l+1\).

As summarized in Table~\ref{tab:integer_transitions}, DH-to-WH requires no basis conversion, DH-to-DH and WH-to-WH require one integer Hadamard transform, and WH-to-DH requires two transforms separated by \(\phi\). Nonhomogeneous activations such as SiLU, GELU~\cite{hendrycks2016gaussian}, and GDN~\cite{balle2015density} instead require explicit fixed-point rescaling on both sides of an integer nonlinear kernel. Supplementary materials give the complete derivation, including bias handling and general nonlinearities. Thus, HaTQ is compatible with integer-only inference.

\begin{figure*}[t]
\centering
\subfloat[Range reduction \label{fig:act_stable}]{%
\includegraphics[width=0.329\textwidth]{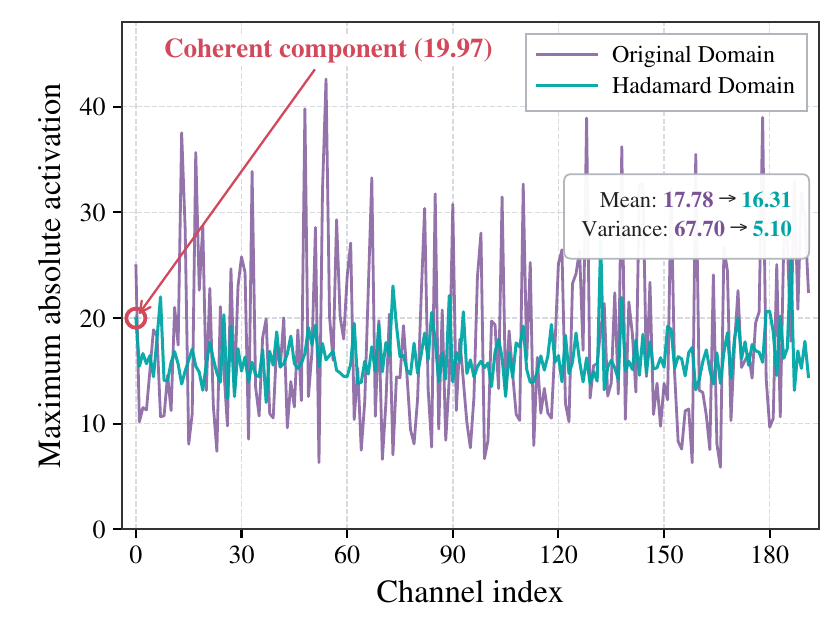}%
}
\subfloat[Coherent amplification \label{fig:act_sensitive_1}]{%
\includegraphics[width=0.329\textwidth]{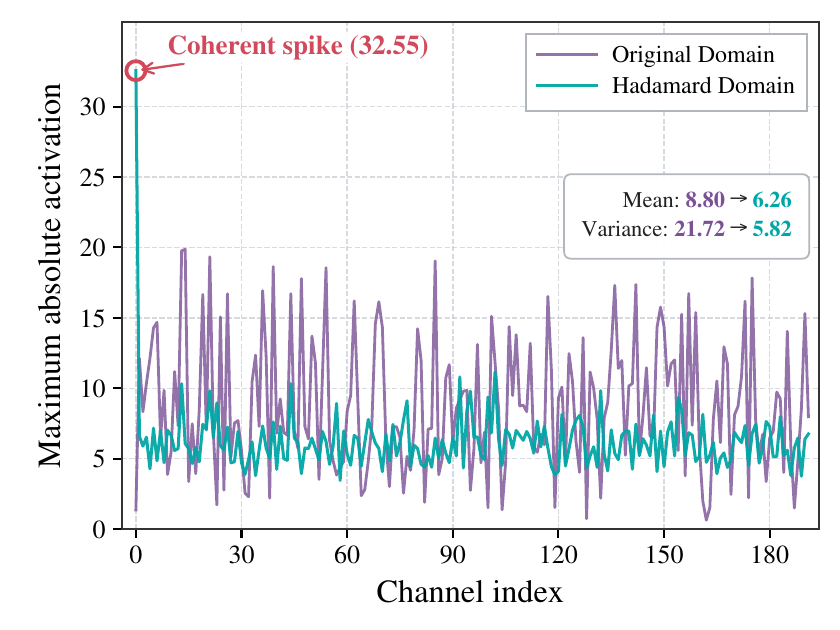}%
}
\subfloat[Coherent amplification \label{fig:act_sensitive_2}]{%
\includegraphics[width=0.329\textwidth]{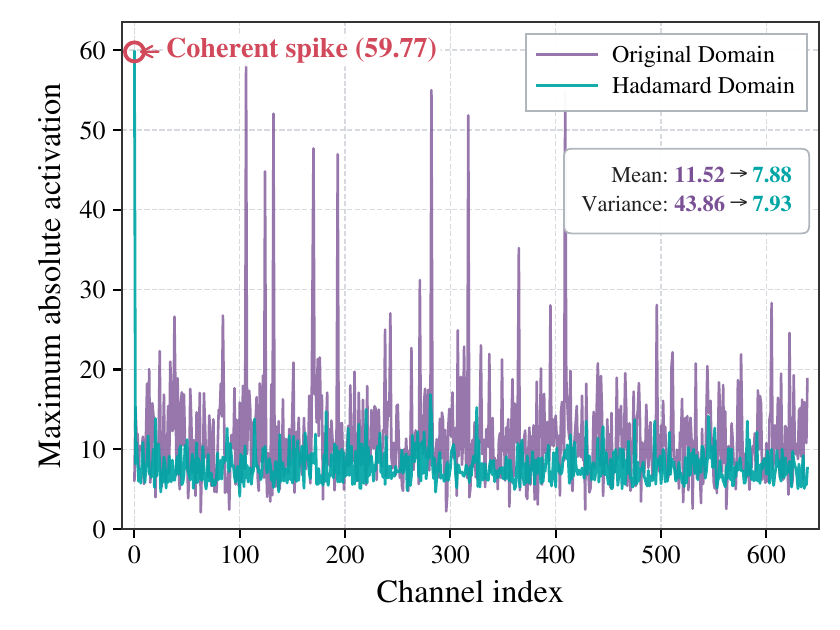}%
}
\caption{Per-channel maximum absolute activations before and after Hadamard transform for three representative neural layers. For most layers, as represented by (a), the constant-basis component remains below the original tensor maximum, and Hadamard transform reduces the overall range. For a small subset of sensitive layers, as exemplified by (b) and (c), coherent accumulation creates new maxima of 32.55 and 59.77, respectively, thereby increasing the scale required by symmetric per-tensor quantization.}
\label{fig:act}
\end{figure*}

\subsection{Layer-wise Hadamard Strategy Selection}
\label{sec:layerwise_hadamard}

Hadamard mixing reduces excess kurtosis for most evaluated tensors. However, we observe that, in a subset of sensitive layers, the maximum activation magnitude increases sharply after the transform. Since the scale of symmetric per-tensor activation quantization is proportional to \(\lVert X_l\rVert_\infty\), such amplification enlarges the quantization range and reduces the effective quantization grid for the remaining values. We therefore select the activation-side transform on a layer-by-layer basis.

\textbf{Activation amplification in sensitive layers.}
To understand why activation-side mixing can be unfavorable, consider the input-channel vector \(\mathbf{x}\in\mathbb{R}^{C_{\mathrm{in},l}}\) at one spatial position of layer \(l\). Let \(m_l^{\mathrm{in}}\geq C_{\mathrm{in},l}\) be the Hadamard order and \(\bar{\mathbf{x}}\in\mathbb{R}^{m_l^{\mathrm{in}}}\) be the corresponding zero-padded vector. We use a normalized Hadamard matrix whose first column is generally defined as the constant basis \(\mathbf{h}_0=\mathbf{1}/\sqrt{m_l^{\mathrm{in}}}\). While the remaining columns combine channels with balanced positive and negative signs, \(\mathbf{h}_0\) adds all channel values with the same sign. Thus, for \(\mathbf{y}=\mathbf{H}_{m_l^{\mathrm{in}}}^{\top}\bar{\mathbf{x}}\), the constant-basis component is
\begin{equation}
y_0
=
\mathbf{h}_0^{\top}\bar{\mathbf{x}}
=
\frac{1}{\sqrt{m_l^{\mathrm{in}}}}
\sum_{j=1}^{C_{\mathrm{in},l}}x_j
=
\frac{C_{\mathrm{in},l}}{\sqrt{m_l^{\mathrm{in}}}}
\mu_{\mathbf{x}},
\label{eq:hadamard_mean_accumulation}
\end{equation}
where \(\mu_{\mathbf{x}}\) is the mean over the original input channels. Eq.~\eqref{eq:hadamard_mean_accumulation} shows that a nonzero channel mean accumulates coherently rather than being canceled by sign mixing. In the unpadded case, \(m_l^{\mathrm{in}}=C_{\mathrm{in},l}\) and \(y_0=\sqrt{C_{\mathrm{in},l}}\,\mu_{\mathbf{x}}\); the constant-basis response is therefore amplified by \(\sqrt{C_{\mathrm{in},l}}\) relative to the channel mean and may become the tensor maximum.

Fig.~\ref{fig:act} illustrates both outcomes. In Fig.~\ref{fig:act}(a), the coherent component remains below the original maximum, so the Hadamard transform reduces the per-tensor range. In Fig.~\ref{fig:act}(b) and (c), it becomes the dominant transformed response and increases the quantization step. Such amplification occurs more often for activations because activation functions can introduce sign asymmetry and nonzero channel means, whereas learned weights are generally more balanced around zero.

\textbf{Profiling-based strategy selection.}
We identify activation-sensitive layers once, before quantization optimization. Given a profiling set of \(M\) images, let \(X_{l,k}\) be the input activation of layer \(l\) for image \(k\), and let \(\mathcal{H}_l^{\mathrm{in}}(\cdot)\) denote its input-side Hadamard transform, including padding when required. We define the empirical range-expansion probability as
\begin{equation}
p_l
=
\frac{1}{M}
\sum_{k=1}^{M}
\mathbb{I}
\left[
\left\lVert
\mathcal{H}_l^{\mathrm{in}}(X_{l,k})
\right\rVert_{\infty}
>
\left\lVert
X_{l,k}
\right\rVert_{\infty}
\right],
\label{eq:hadamard_sensitivity}
\end{equation}
where \(\mathbb{I}[\cdot]\) is the indicator function. A large $p_l$ indicates that activation-side Hadamard transformation consistently enlarges the quantization range of layer $l$.

Based on this statistic, the layer-wise strategy is selected as
\begin{equation}
\pi_l =
\begin{cases}
\mathrm{WH}, & p_l > \tau_{\mathrm{sel}},\\
\mathrm{DH}, & p_l \leq \tau_{\mathrm{sel}},
\end{cases}
\label{eq:layerwise_hadamard_selection}
\end{equation}
where DH and WH are defined in Sec.~\ref{sec:hadamard_quantization}. Layers with frequent range expansion use WH to avoid transforming their activations; all remaining layers use DH.

We use \(M=1000\) profiling images and \(\tau_{\mathrm{sel}}=0.8\) in all experiments. The resulting \(\pi_l\) is fixed during PTQ, QAT, and inference, introducing neither input-dependent branching nor online selection overhead.

\subsection{Hadamard Construction and Implementation}
\label{sec:hadamard_construction}
Hadamard matrices are not guaranteed to exist for arbitrary orders. For a channel dimension \(C\), we therefore pad it to the smallest supported order no smaller than the next multiple of four:
\begin{equation}
C_4(C)=4\left\lceil\frac{C}{4}\right\rceil,
\quad
m(C)
=
\min\left\{m\in\mathcal{M}_{H}\mid m\geq C_4(C)\right\},
\label{eq:hadamard_order}
\end{equation}
where \(\mathcal{M}_{H}\) contains the orders supported by the adopted Sylvester, Paley, and Kronecker constructions~\cite{seberry2020hadamard, hedayat1978hadamard}. The constructions are detailed in supplementary materials.

The transformed weights are input-independent and can therefore be precomputed offline. During inference, each activation-side Hadamard transform is implemented as a fixed bias-free ($1\times1$) convolution, ensuring compatibility with standard inference backends. This portable implementation is not computationally optimal, as it realizes the signed transform as a dense convolution, incurs redundant multiply--accumulate and memory operations, and does not exploit operator fusion. The overhead could be further reduced through dedicated addition--subtraction kernels and fusion with adjacent quantized operators. Since these optimizations require backend- and hardware-specific CUDA kernel design, we leave them for future work.

\subsection{Training and Optimization}
\label{sec:whole_network_optimization}

Our framework supports two quantization workflows, QAT and PTQ, as well as two quantization granularities, channel-wise and tensor-wise quantization. These two choices can be combined independently, resulting in four supported configurations. After determining the configuration and fixing the transform type for each layer, the complete quantized codec is trained or calibrated accordingly.

\textbf{Network-wise PTQ calibration.}
Conventional PTQ methods commonly calibrate individual layers or blocks by minimizing local reconstruction error, as in AdaRound~\cite{adaround}, BRECQ~\cite{brecq} and RDO-PTQ~\cite{shi2023rate}. Such local objectives, however, cannot fully capture how quantization errors propagate across the codec. Following the network-wise paradigm of NeuroQuant~\cite{neuroquant}, we instead optimize all learnable quantization parameters jointly using
\begin{equation}
\mathcal{L}_{\mathrm{PTQ}}=R+\lambda D+\eta\mathcal{L}_{\mathrm{reg}},
\label{eq:network_wise_ptq}
\end{equation}
where \(R\) is the estimated rate, \(D\) is the reconstruction distortion, \(\lambda\) controls their trade-off, and \(\eta\) weights the adaptive-rounding regularizer \(\mathcal{L}_{\mathrm{reg}}\). The regularizer progressively drives the learnable rounding variables toward discrete decisions.

For channel-wise activation quantization, DH is adopted for all layers, whereas tensor-wise activation quantization uses the calibration set to select between DH and WH for each layer according to Eq.~\eqref{eq:layerwise_hadamard_selection}. The activation quantization ranges are estimated dynamically for each input at inference time.

\textbf{Quantization-aware training.}
For QAT, we fine-tune the network weights and quantization parameters using
\begin{equation}
\mathcal{L}_{\mathrm{QAT}}=R+\lambda D.
\label{eq:qat_objective}
\end{equation}
The layer strategies remain fixed, while the activation scales are tracked during training and frozen for static inference.

Let \(X_t\) denote the activation presented to a quantizer at step \(t\): it is expressed in the input Hadamard basis for DH and in the original basis for WH. For tensor-wise symmetric \(b\)-bit quantization, the instantaneous scale is
\begin{equation}
\widetilde{s}_t
=
\frac{\lVert X_t\rVert_\infty}{q_{\max}^{\mathrm{s}}},
\qquad
q_{\max}^{\mathrm{s}}=2^{b-1}-1.
\label{eq:current_scale}
\end{equation}

For channel-wise quantization, \(\widetilde{s}_t\) and the running scale \(s_t\) are vectors, and all subsequent operations are applied independently to each channel. Because an isolated spike can corrupt a conventional exponential moving average (EMA), we update the running scale only when \(\widetilde{s}_t\) is within a prescribed multiple of its previous value:
\begin{equation}
s_t=
\begin{cases}
\beta_{\mathrm{ema}}s_{t-1}
+(1-\beta_{\mathrm{ema}})\widetilde{s}_t,
& \widetilde{s}_t\leq\tau_{\mathrm{ema}}s_{t-1},\\[1mm]
s_{t-1},
& \widetilde{s}_t>\tau_{\mathrm{ema}}s_{t-1},
\end{cases}
\label{eq:robust_ema_scale}
\end{equation}
with \(s_0=\widetilde{s}_0\). We use \(\beta_{\mathrm{ema}}=0.99\) and \(\tau_{\mathrm{ema}}=15\). The final \(s_t\) values are fixed for inference.

\begin{figure*}[!t]
\centering
\subfloat[Kodak\label{fig:rd_kodak}]{%
\includegraphics[width=0.329\textwidth]{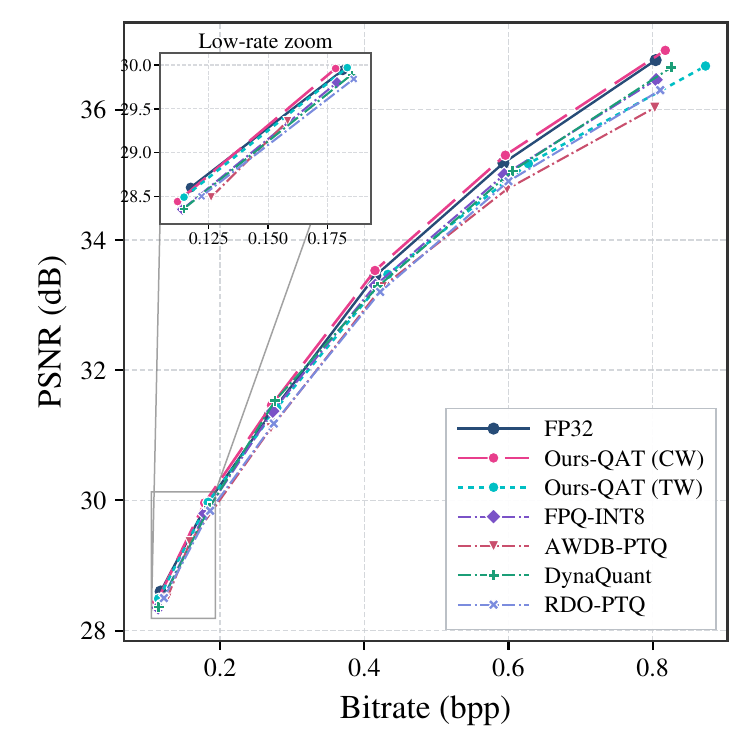}%
}\hspace{0.003\textwidth}%
\subfloat[CLIC\label{fig:rd_clic}]{%
\includegraphics[width=0.329\textwidth]{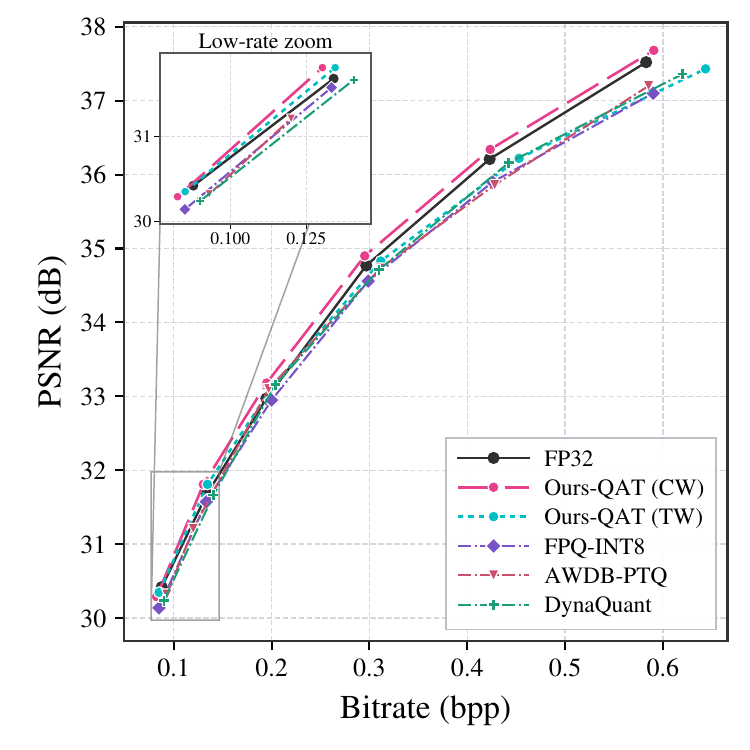}%
}\hspace{0.003\textwidth}%
\subfloat[Tecnick\label{fig:rd_tecnick}]{%
\includegraphics[width=0.329\textwidth]{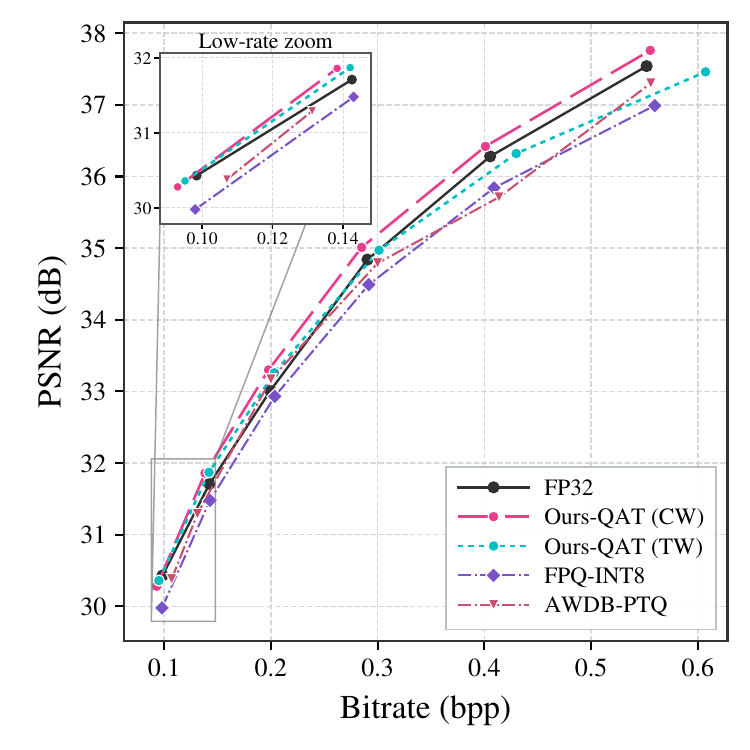}%
}
\caption{Rate--distortion performance of different quantization methods applied to GMM~\cite{cheng2020learned} on the Kodak, CLIC, and Tecnick datasets. The original FP32 model serves as the anchor. CW and TW denote channel-wise and tensor-wise activation quantization, respectively.}
\label{fig:rd}
\end{figure*}

\section{Experiment}

\subsection{Experimental Setup}

\textbf{Models and operating points.}
We evaluate the proposed method on two representative LIC architectures, GMM~\cite{cheng2020learned} and ELIC~\cite{he2022elic}. Their available pretrained checkpoints\footnote{\url{https://interdigitalinc.github.io/CompressAI/zoo.html}}\footnote{\url{https://github.com/VincentChandelier/ELiC-ReImplemetation}} serve as both the quantization initialization and the corresponding FP32 references. For each architecture, we evaluate six fixed-rate models. The R--D multipliers are $\lambda\in\{0.0018,0.0035,0.0067,0.013,0.025,0.0483\}$ for GMM and $\lambda\in\{0.0004,0.0008,0.0016,0.0032,0.015,0.045\}$ for ELIC, following their open-source configurations.

\textbf{Training, calibration, and profiling.}
QAT uses MLIC-Train-100K~\cite{jiang2023mlic} and fine-tunes each pretrained model for 50 epochs with a learning rate of $1\times10^{-5}$. PTQ uses the 512 images from DIV2K~\cite{Agustsson_2017_CVPR_Workshops} as the calibration set. The layer-wise DH/WH assignment is determined before quantization optimization using $M=1000$ profiling images and the range-expansion statistic in Eq.~\eqref{eq:hadamard_sensitivity}. We set $\tau_{\mathrm{sel}}=0.8$ according to the empirical range-expansion statistics, so that WH is assigned only to layers exhibiting persistent activation-range amplification. The resulting assignment is fixed throughout PTQ, QAT, and inference. All QAT and PTQ optimization runs are conducted on a single NVIDIA RTX 4090D GPU.

\textbf{Evaluation datasets and metrics.}
We evaluate all models on Kodak~\cite{kodak1993kodak}, Tecnick~\cite{asuni2014testimages}, and the CLIC Professional validation set~\cite{toderici2020workshop}. Kodak contains 24 images at $512\times768$, Tecnick contains 100 images at $1200\times1200$, and CLIC contains 41 2K-resolution images. Rate--distortion performance is measured using bits per pixel (bpp) and peak signal-to-noise ratio (PSNR). We further report the Bj{\o}ntegaard delta rate (BD-rate)~\cite{bjontegaard2001calculation} relative to the corresponding FP32 checkpoints over the six operating points. Lower BD-rate values indicate better performance, and negative values indicate an improvement over the FP32 reference.

\begin{table*}[t]
\centering
\caption{BD-rate change (\%) of quantized GMM models relative to their FP32 counterparts.}
\label{tab:cheng_quantization_comparison}
\footnotesize
\setlength{\tabcolsep}{4.5pt}
\renewcommand{\arraystretch}{1.12}
\begin{tabular*}{\textwidth}{@{\extracolsep{\fill}}llllcccc@{}}
\toprule
\textbf{Granularity}
& \textbf{Optimization}
& \textbf{Method}
& \textbf{Precision}
& \textbf{Kodak}
& \textbf{CLIC}
& \textbf{Tecnick}
& \textbf{Avg.}
\\
\midrule

\multirow{6}{*}{Channel-wise}
& \multirow{6}{*}{QAT}
& DynaQuant (AAAI'26)~\cite{bao2026dynaquant}
& UP-INT8
& $+1.02\%$
& $+2.61\%$
& --
& $+1.82\%$
\\
&
& DynaQuant (AAAI'26)~\cite{bao2026dynaquant}
& MP-INT6.2
& $+7.15\%$
& $+12.87\%$
& --
& --
\\
&
& FPQ (ICME'24)~\cite{hossain2024flexible}
& UP-INT8
& $+2.05\%$
& $+3.54\%$
& $+4.97\%$
& $+3.52\%$
\\
&
& FMPQ (ICME'24)~\cite{hossain2024flexible}
& MP-INT8.0
& $+0.89\%$
& $+1.70\%$
& $+2.68\%$
& $+1.76\%$
\\
&
& Q-LIC (TCSVT'25)~\cite{sun2022q}
& UP-INT8
& $+10.50\%$
& $+13.00\%$
& --
& $+11.75\%$
\\
&
& \textbf{HaTQ (Ours)}
& UP-INT8
& $\best{-1.84\%}$
& $\best{-4.18\%}$
& $\best{-5.55\%}$
& $\best{-3.86\%}$
\\
\midrule

\multirow{3}{*}{Channel-wise}
& \multirow{3}{*}{PTQ}
& RDO-PTQ (TCSVT'24)~\cite{shi2023rate}
& UP-INT8
& $+4.88\%$
& --
& $+6.86\%$
& $+5.87\%$
\\
&
& SS-PTQ (IEEE Access'25)~\cite{yang2025subset}
& NU-INT8
& $+4.05\%$
& --
& --
& $+4.05\%$
\\
&
& \textbf{HaTQ (Ours)}
& UP-INT8
& $\best{+1.13\%}$
& $\best{+1.26\%}$
& $\best{+2.17\%}$
& $\best{+1.52\%}$
\\
\midrule

\multirow{2}{*}{Tensor-wise}
& \multirow{2}{*}{QAT}
& RAQ-QAT (TCSVT'21)~\cite{hong2020efficient}
& UP-INT8
& $+12.20\%$
& $+17.60\%$
& --
& $+14.90\%$
\\
&
& \textbf{HaTQ (Ours)}
& UP-INT8
& $\best{+4.29\%}$
& $\best{+1.98\%}$
& $\best{-0.08\%}$
& $\best{+2.06\%}$
\\
\midrule

\multirow{7}{*}{Tensor-wise}
& \multirow{7}{*}{PTQ}
& RAQ-PTQ (TCSVT'21)~\cite{hong2020efficient}
& UP-INT8
& $+27.84\%$
& --
& $+29.95\%$
& $+28.90\%$
\\
&
& AWDB-PTQ (ACM MM'25)~\cite{yu2025activation}
& UP-INT8
& $+8.74\%$
& $+13.06\%$
& $+15.53\%$
& $+12.44\%$
\\
&
& LinearQuant (ACM MM'25)~\cite{yu2025activation}
& UP-INT9
& $+5.51\%$
& $+7.29\%$
& $+8.47\%$
& $+7.09\%$
\\
&
& SmoothQuant (ICML'23)~\cite{xiao2023smoothquant}
& UP-INT9
& $\best{+2.35\%}$
& $+3.92\%$
& $+4.41\%$
& $+3.56\%$
\\
&
& AWDB-PTQ (ACM MM'25)~\cite{yu2025activation}
& UP-INT9
& $+2.85\%$
& $+3.77\%$
& $+5.57\%$
& $+4.06\%$
\\
&
& MPPTQ (IoT-J'25)~\cite{yu2025mixed}
& MP-INT11.3$^{\dagger}$
& $+6.29\%$
& $+8.36\%$
& $+9.57\%$
& $+8.07\%$
\\
&
& \textbf{HaTQ (Ours)}
& UP-INT8
& $+3.15\%$
& $\best{+2.95\%}$
& $\best{+2.77\%}$
& $\best{+2.96\%}$
\\
\bottomrule
\end{tabular*}
\vspace{1mm}

\parbox{\textwidth}{\footnotesize \textit{Note:} UP, NU, and MP denote uniform, non-uniform, and mixed-precision quantization, respectively. INT8 and INT9 indicate 8- and 9-bit weight--activation configurations. For MP, precision denotes the approximate equivalent bitwidth. MPPTQ uses layer-wise configurations selected from W4A4, W8A8, and W16A16. Layer-wise and tensor-wise activation quantization are equivalent and differ only in terminology. Averages are calculated over the available datasets. Best results within each setting are $\best{highlighted}$.}
\end{table*}

\begin{table*}[t]
\centering
\caption{BD-rate change (\%) of quantized ELIC models relative to their FP32 counterparts.}
\label{tab:elic_quantization_comparison}
\footnotesize
\setlength{\tabcolsep}{4.5pt}
\renewcommand{\arraystretch}{1.12}
\begin{tabular*}{\textwidth}{@{\extracolsep{\fill}}llllcccc@{}}
\toprule
\textbf{Granularity}
& \textbf{Optimization}
& \textbf{Method}
& \textbf{Precision}
& \textbf{Kodak}
& \textbf{CLIC}
& \textbf{Tecnick}
& \textbf{Avg.}
\\
\midrule

\multirow{3}{*}{Channel-wise}
& \multirow{3}{*}{QAT}
& DynaQuant (AAAI'26)~\cite{bao2026dynaquant}
& UP-INT8
& $+5.97\%$
& $+2.55\%$
& --
& $+4.26\%$
\\
&
& DynaQuant (AAAI'26)~\cite{bao2026dynaquant}
& MP-INT6.9
& $+7.62\%$
& $+4.01\%$
& --
& --
\\
&
& \textbf{HaTQ (Ours)}
& UP-INT8
& $\best{+0.16\%}$
& $\best{+0.63\%}$
& $\best{+6.33\%}$
& $\best{+2.37\%}$
\\
\midrule

Tensor-wise
& QAT
& \textbf{HaTQ (Ours)}
& UP-INT8
& $\best{+5.03\%}$
& $\best{+6.40\%}$
& $\best{+8.33\%}$
& $\best{+6.59\%}$
\\
\bottomrule
\end{tabular*}
\vspace{1mm}

\parbox{\textwidth}{\footnotesize \textit{Note:} UP and MP denote uniform- and mixed-precision quantization, respectively. INT8 indicates an 8-bit weight--activation configuration. For MP, precision denotes the approximate equivalent bitwidth. Averages are calculated over the available datasets. Best results within each setting are $\best{highlighted}$.}
\end{table*}
\textbf{Comparison methods.}
We compare against representative LIC quantization methods under their reported optimization regimes and activation granularities. The QAT baselines comprise Q-LIC~\cite{sun2022q}, RAQ-QAT~\cite{hong2020efficient}, DynaQuant~\cite{bao2026dynaquant}, FPQ~\cite{hossain2024flexible}, and FMPQ~\cite{hossain2024flexible}. The PTQ baselines comprise RDO-PTQ~\cite{shi2023rate}, SS-PTQ~\cite{yang2025subset}, RAQ-PTQ~\cite{hong2020efficient}, MPPTQ~\cite{yu2025mixed}, and AWDB-PTQ~\cite{yu2025activation}. We additionally include LinearQuant~\cite{nagel2021white} and SmoothQuant~\cite{xiao2023smoothquant} using the results reported by Yu et al.~\cite{yu2025activation}. These baselines cover uniform, non-uniform, and mixed-precision quantization under both channel-wise and tensor-wise activation granularities. Tables~\ref{tab:cheng_quantization_comparison} and~\ref{tab:elic_quantization_comparison} report the exact precision, quantization granularity, optimization regime, and benchmark results for each comparison.

\textbf{Quantization configurations.}
Unless otherwise specified, every HaTQ variant uses uniform W8A8 quantization. We apply channel-wise symmetric quantization to weights and evaluate both channel-wise and tensor-wise activation quantization. PTQ uses static weight quantization and dynamic affine activation quantization. QAT uses static symmetric quantization, with activation scales tracked during fine-tuning and frozen before inference. The tensor-wise QAT configuration follows the regular W8A8 execution pattern supported by standard INT8 inference engines. We deploy this configuration with NVIDIA TensorRT and report its runtime overhead in the deployment study.

\begin{figure*}[!t]
\centering
\subfloat[GMM, Kodak\label{fig:cheng_kodak}]{%
\includegraphics[width=0.329\textwidth]{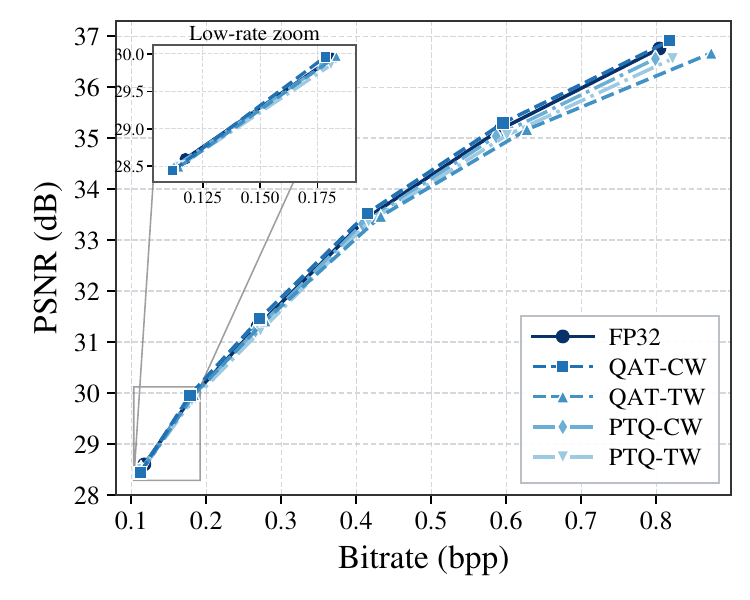}%
}
\subfloat[GMM, CLIC\label{fig:cheng_clic}]{%
\includegraphics[width=0.329\textwidth]{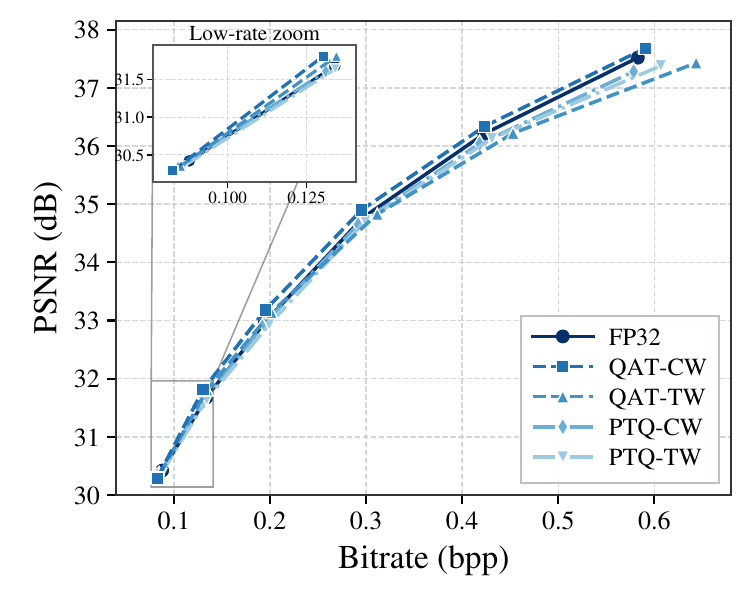}%
}
\subfloat[GMM, Tecnick\label{fig:cheng_tecnick}]{%
\includegraphics[width=0.329\textwidth]{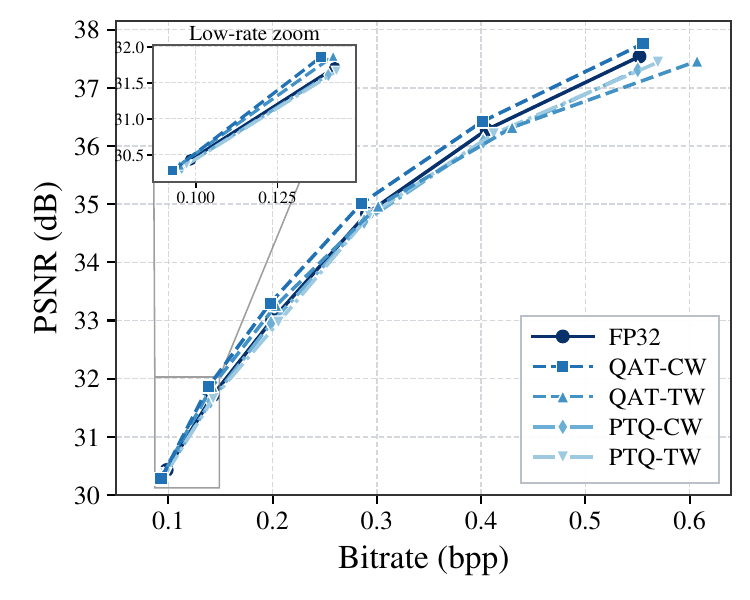}%
}\\
\subfloat[ELIC, Kodak\label{fig:elic_kodak}]{%
\includegraphics[width=0.329\textwidth]{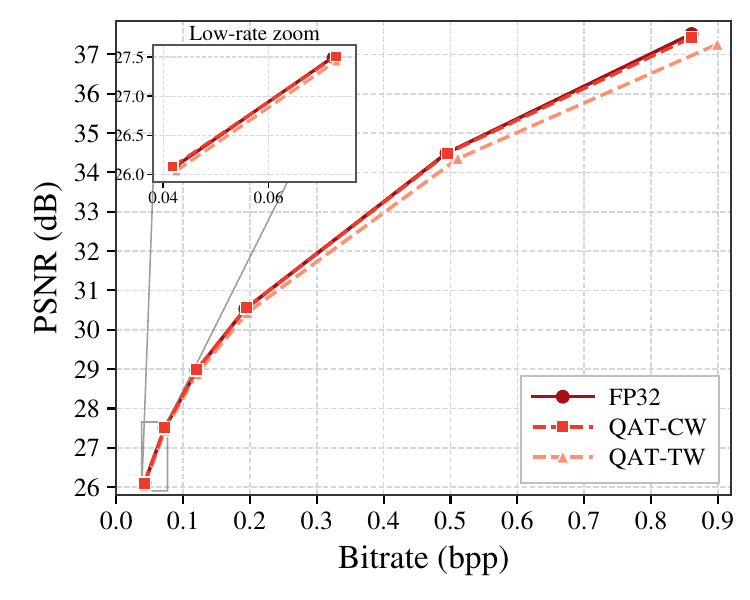}%
}
\subfloat[ELIC, CLIC\label{fig:elic_clic}]{%
\includegraphics[width=0.329\textwidth]{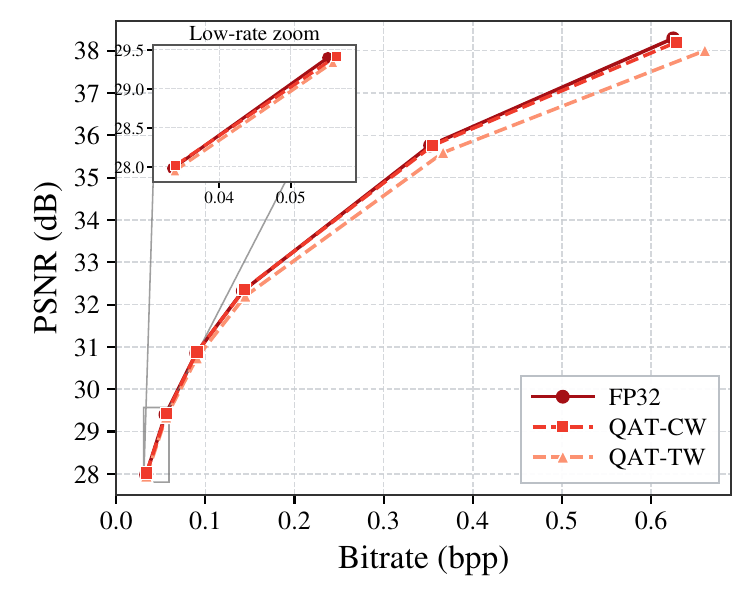}%
}
\subfloat[ELIC, Tecnick\label{fig:elic_tecnick}]{%
\includegraphics[width=0.329\textwidth]{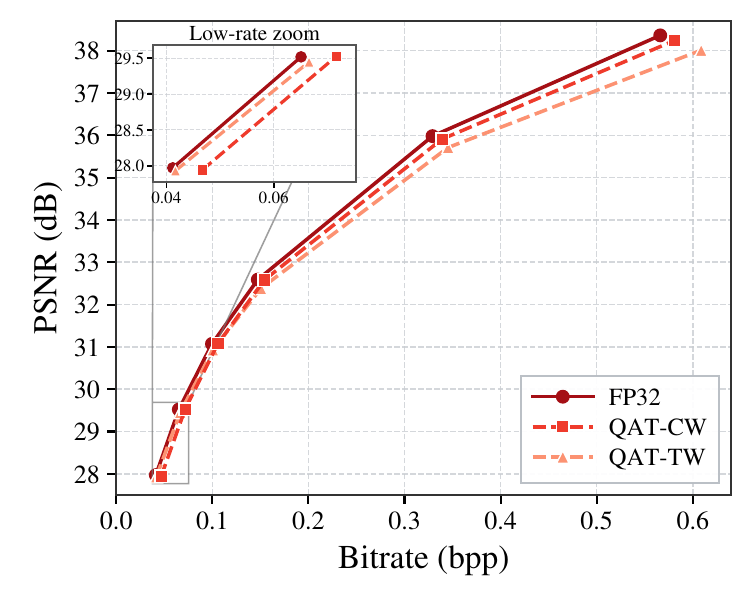}%
}
\caption{Rate--distortion performance of the FP32 and INT8 GMM (top) and ELIC (bottom) under different quantization settings. CW and TW denote channel-wise and tensor-wise activation quantization, respectively.}
\label{fig:rd_comparison}
\end{figure*}

\subsection{Evaluation}

Fig.~\ref{fig:rd} and~\ref{fig:rd_comparison} present the R--D curves of the FP32 and INT8 models, while Tables~\ref{tab:cheng_quantization_comparison} and~\ref{tab:elic_quantization_comparison} report the corresponding BD-rate changes.

\textbf{Comparison with QAT methods.}
Existing state-of-the-art QAT methods mainly address heterogeneous quantization sensitivity by increasing the flexibility of the quantizer. FMPQ~\cite{hossain2024flexible} assigns mixed precision to sensitive layers, whereas DynaQuant~\cite{bao2026dynaquant} uses content-adaptive quantization bitwidth. In contrast, HaTQ retains uniform W8A8 quantization. On GMM, channel-wise HaTQ-QAT achieves an average BD-rate change of $-3.86\%$, outperforming both mixed-precision FMPQ and DynaQuant. The R--D curves in Fig.~\ref{fig:rd} show that this advantage is maintained across the six operating points. Redistributing concentrated responses can therefore be more effective than assigning additional precision to selected layers. The gains also transfer to ELIC. On Kodak, channel-wise HaTQ-QAT reduces the BD-rate changes of mixed-precision DynaQuant from $5.97\%$ to $0.16\%$. The improvement across two different LIC architectures suggests that the benefit is not specific to the transform and context structure of GMM. 

Notably, HaTQ-QAT achieves negative BD-rate values relative to the corresponding FP32 checkpoints. This does not imply that INT8 arithmetic is intrinsically superior to FP32, as the pretrained FP32 checkpoints are not strict R--D upper bounds. A plausible explanation is that Hadamard-domain reparameterization improves QAT conditioning, enabling convergence to more favorable R--D operating points. How parameterization affects the loss landscape and training dynamics represents another interesting direction for future research.

\textbf{Comparison with PTQ methods.}
Across the Kodak and Tecnick benchmarks, HaTQ reduces the average BD-rate increase from \(5.87\%\) for RDO-PTQ to \(1.65\%\). Unlike RDO-PTQ~\cite{shi2023rate}, which optimizes individual layers independently, our network-wise calibration coordinates quantization errors according to their accumulated impact on the end-to-end R--D objective. SS-PTQ~\cite{yang2025subset} enhances weight representability through non-uniform quantization but operates entirely in the original weight domain. On Kodak, HaTQ achieves a BD-rate increase of only \(1.13\%\), compared with \(4.05\%\) for SS-PTQ. This result demonstrates that uniform quantization in a suitably conditioned transform domain can outperform more flexible non-uniform quantization applied directly in the original domain.

\textbf{Performance under tensor-wise quantization.}
The benefit of HaTQ is retained when all activation channels share one quantization scale. This setting is more sensitive to channel-wise range imbalance because a few extreme responses can determine the quantization grid of the entire tensor. Under tensor-wise PTQ, HaTQ reduces the average BD-rate degradation of AWDB-PTQ~\cite{yu2025activation} from $12.44\%$ to $2.96\%$. AWDB-PTQ is a channel-wise rescaling method that balances the activation and weight ranges by transferring the dynamic-range burden between the two operands. In contrast, the Hadamard transform redistributes concentrated components within the transform domain through orthogonal mixing.

Overall, HaTQ consistently improves both QAT and PTQ, remains effective with tensor-wise activation quantization, and transfers from GMM to ELIC. The results support transform-domain quantization as a practical alternative to mixed-precision or non-uniform quantization for INT8 LIC deployment.

\begin{figure*}[!t]
\centering
\subfloat[Channel-wise PTQ\label{fig:ablation_channel_ptq}]{%
\includegraphics[width=0.246\textwidth]{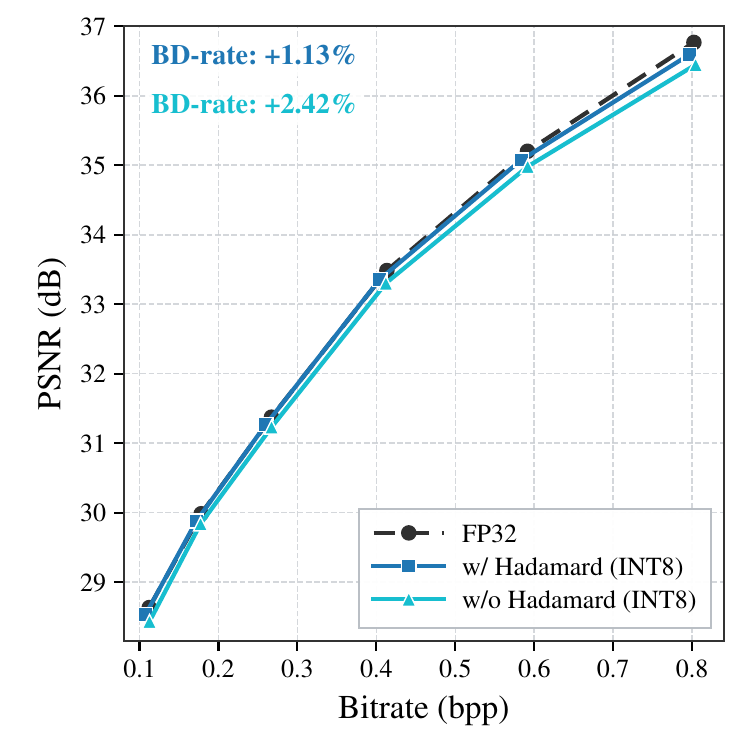}%
}
\subfloat[Channel-wise QAT\label{fig:ablation_channel_qat}]{%
\includegraphics[width=0.246\textwidth]{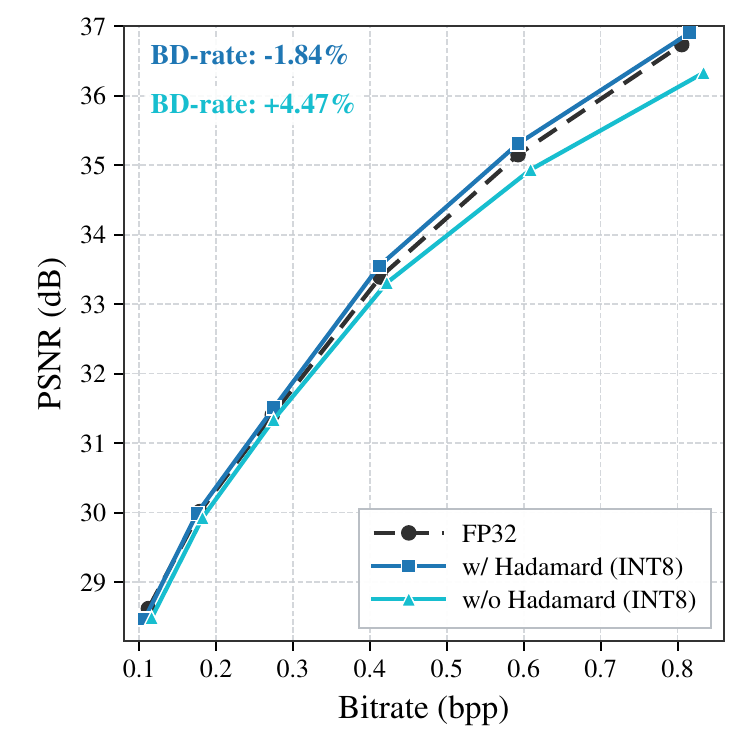}%
}
\subfloat[Tensor-wise PTQ\label{fig:ablation_tensor_ptq}]{%
\includegraphics[width=0.246\textwidth]{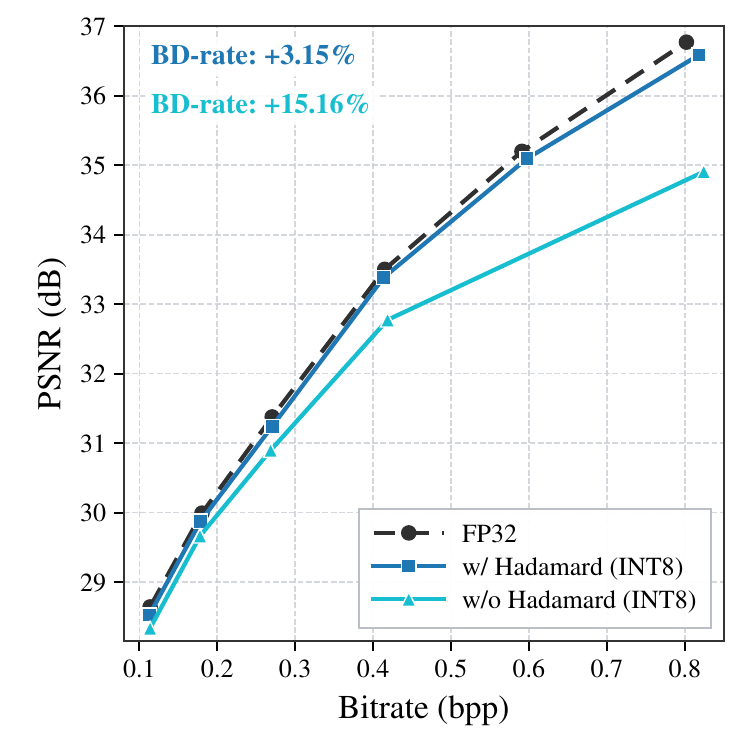}%
}
\subfloat[Tensor-wise QAT\label{fig:ablation_tensor_qat}]{%
\includegraphics[width=0.246\textwidth]{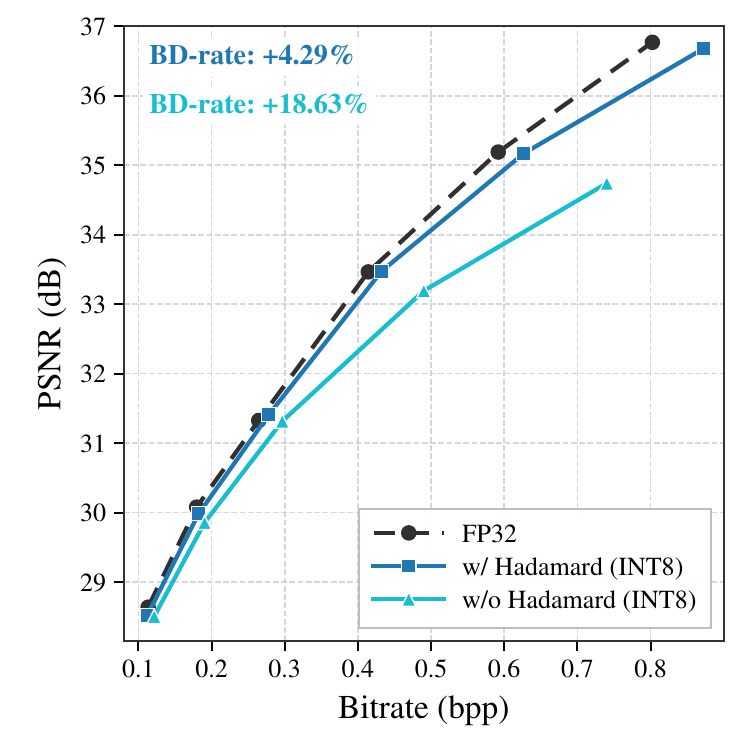}%
}
\caption{Effectiveness of the Hadamard transform across different quantization settings.}
\label{fig:hadamard_ablation}
\end{figure*}

\subsection{Ablation Studies}

Unless otherwise specified, all ablation studies are conducted on Kodak using GMM.

\textbf{Effect of Hadamard-domain quantization.}
We first remove all Hadamard transforms while retaining the quantizers, training data, and optimization schedule. Across the four settings in Fig.~\ref{fig:hadamard_ablation}, the transformed models consistently outperform their counterparts. Thus, the improvement demonstrates that the domain presented to the uniform quantizer is itself a principal source of error.

The magnitude of the gain reveals when this transformation is most useful. With channel-wise activation scales, PTQ with dynamic activation quantization can already absorb much of the inter-channel range variation on each input. Static QAT scales cannot adapt in this manner and therefore benefit more from Hadamard reparameterization. The distinction becomes sharper under tensor-wise activation quantization. In this case, some extreme components determine the step size for the complete tensor. The agreement between the distributional and R--D changes supports Hadamard-domain quantization.

\begin{figure*}[t]
\centering
\subfloat[Layer-adaptive selection mechanism.\label{fig:las}]{%
    \includegraphics[width=0.32\textwidth]{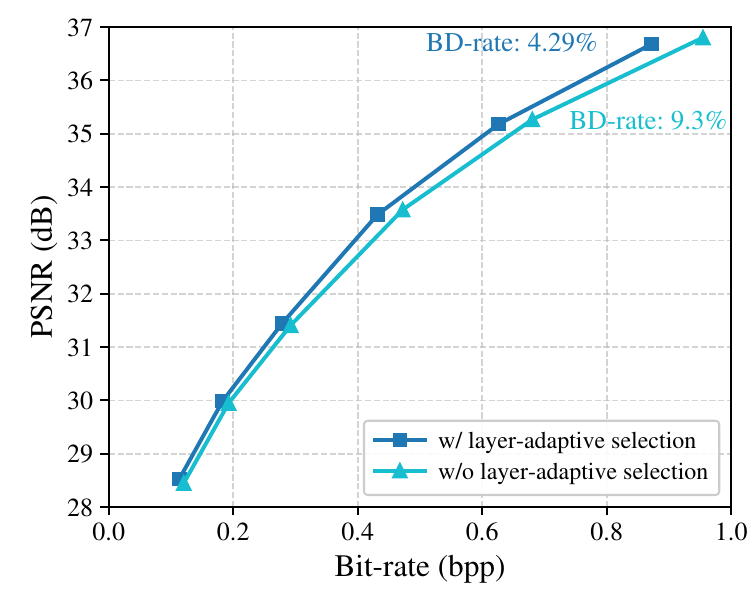}%
}
\hfill
\subfloat[Calibration set size.\label{fig:calib}]{%
    \includegraphics[width=0.32\textwidth]{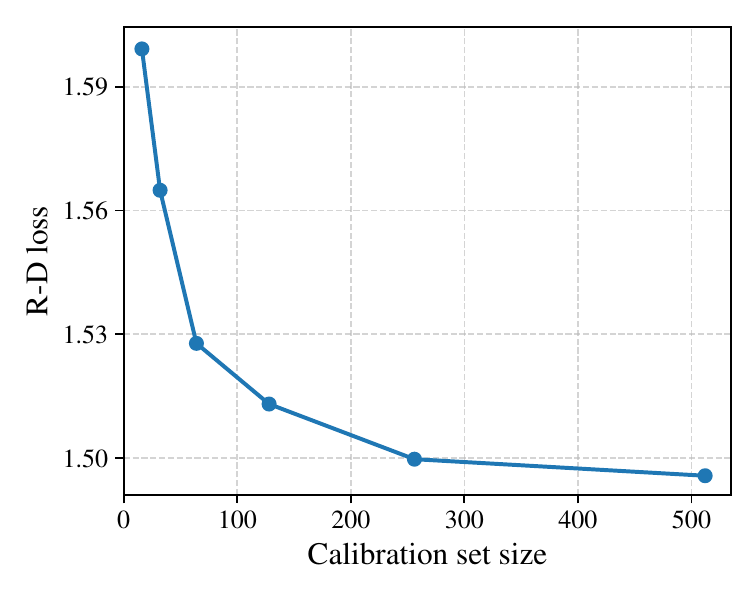}%
}
\hfill
\subfloat[Inference latency.\label{fig:latency}]{%
    \includegraphics[width=0.32\textwidth]{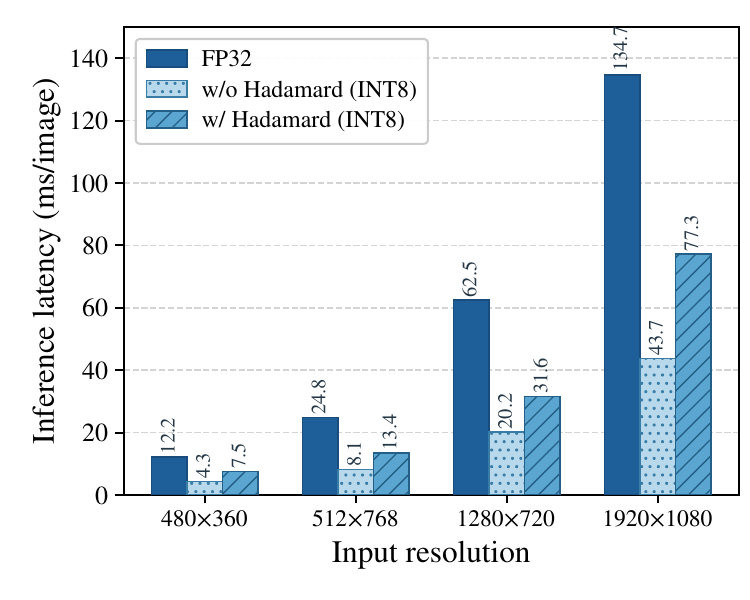}%
}
\caption{Ablation studies and deployment analysis. (a) Effect of layer-adaptive selection on the R--D performance of GMM on Kodak under tensor-wise activation QAT. (b) Effect of calibration set size on whole-network R--D optimization. (c) Inference latency of FP32 and INT8 models with and without the Hadamard transform at different input resolutions.}
\label{fig:ablation_and_deployment}
\end{figure*}

\textbf{Effect of layer-adaptive selection.}
In layers with a nonzero channel mean, the constant Hadamard basis can coherently accumulate activation components and increase this maximum even when the overall distribution becomes less heavy-tailed. Profiling shows that the sensitive layers are mainly located in the context module, where quantization errors directly perturb probability estimation and accumulate over latent symbols. Our profiling criterion in Eq.~\eqref{eq:hadamard_sensitivity} therefore applies double-Hadamard quantization to beneficial layers and switches sensitive layers to weight-only transformation. As shown in Fig.~\ref{fig:las}, this reduces the BD-rate loss from 9.30\% to 4.29\%, demonstrating that layer adaptivity is necessary to prevent locally unfavorable transformations from degrading the end-to-end coding objective.


\textbf{Effect of calibration-set size.} As shown in Fig.~\ref{fig:calib}, the R--D loss decreases steadily and approaches saturation at approximately 512 images. Unlike layer-wise reconstruction, whole-network calibration jointly optimizes quantization parameters according to their accumulated effects on entropy estimation and reconstruction, and therefore benefits from greater data diversity. We consequently use 512 calibration images in PTQ.

\subsection{Deployment}
\label{sec:deployment}

We evaluate the deployment characteristics of HaTQ on NVIDIA TensorRT. Both HaTQ and the baseline use uniform W8A8 quantization with tensor-wise activation scales and channel-wise weight scales.

\textbf{TensorRT realization.}
As described in Sec.~\ref{sec:hadamard_construction}, all weight-side transforms are folded offline into the padded convolution kernels and therefore introduce no additional runtime operators. Each online Hadamard transform is exported as a fixed, bias-free \(1\times1\) convolution whose coefficients are specified by the corresponding Hadamard matrix. The remaining network is executed using the standard TensorRT INT8 pipeline. This implementation requires no custom TensorRT plugins and supports both the power-of-two and non-power-of-two Hadamard orders used in our construction.

\textbf{Latency and storage overhead.}
We measure average inference latency on an NVIDIA GeForce RTX 4090D at the different input resolutions shown in Fig.~\ref{fig:latency}. At \(768\times512\), HaTQ requires \(13.39\) ms per image, compared with \(8.13\) ms for the TensorRT INT8 QAT baseline and \(24.80\) ms for the FP32 reference. Thus, the portable HaTQ realization incurs \(5.26\) ms over the matched INT8 baseline while remaining \(11.41\) ms faster than the reported floating-point implementation. The Hadamard matrices add only \(1.5\) MB of storage.




\textbf{Optimization potential.}
The measured overhead primarily reflects the portable realization of the online transform. Treating a signed Hadamard transform as an unfused dense \(1\times1\) convolution introduces avoidable multiplications, memory traffic, and kernel launches. As derived in Sec.~\ref{sec:integer_only_inference}, the transform itself requires only additions and subtractions, and its normalization factor can be absorbed into adjacent fixed-point scales. Dedicated integer Hadamard kernels, fusion with neighboring quantized operators, and butterfly implementations for power-of-two orders can therefore reduce the deployment overhead without changing the trained model.

\section{Conclusion}
This work reframes LIC quantization as a problem of selecting a better numerical domain, rather than continually increasing the flexibility of the quantizer. We introduced equivalent Hadamard reparameterizations that redistribute outliers and channel-wise magnitude variations before uniform INT8 quantization. The proposed framework consistently improves R--D performance under all four combinations of quantization workflow and granularity across different LIC architectures and datasets. More broadly, the results connect model quantization with the principle of transform coding: a simple quantizer can become more effective when its input representation is better conditioned. The operators are compatible with integer-only execution, and the TensorRT implementation demonstrates that the proposed method can be readily integrated into practical inference pipelines. The current implementation uses unfused \(1\times1\) convolutions for online transforms and simplifies some branches, leaving clear room for improvement. Dedicated Hadamard kernels, operator fusion, and extensions to lower bit widths are promising directions for future work.


\bibliographystyle{IEEEtran}
\bibliography{reference}


 





\clearpage
\newpage

\twocolumn[{%
\begin{minipage}{\textwidth}
\centering

{\Large\bfseries Supplementary Material\par}
\vspace{0.5em}

{\Large Hadamard-Domain Model Quantization for Learned Image Coding\par}
\vspace{2em}

\end{minipage}%
}]

This supplementary material provides the Hadamard constructions, the
function-preserving padded forms of double-Hadamard (DH) and weight-only
Hadamard (WH) quantization, and the integer realization for bias terms and
general nonlinearities. The notation follows the main paper. In particular, a
linear or unfolded convolutional layer is written as
\begin{equation}
Y_l=X_lW_l^{\top}+b_l,
\qquad
X_l\in\mathbb R^{N_l\times D_l},
\quad
W_l\in\mathbb R^{C_{\mathrm{out},l}\times D_l},
\label{eq:supp_linear_operator}
\end{equation}
where the bias \(b_l\) is broadcast over the \(N_l\) rows.

\section{Hadamard Construction and Padded Equivalence}
\label{sec:supp_hadamard}

\subsection{Supported Hadamard Orders}

An unnormalized Hadamard matrix
\(\widetilde{\mathbf H}_m\in\{-1,+1\}^{m\times m}\) and its normalized form
\(\mathbf H_m\) satisfy
\begin{align}
&\widetilde{\mathbf H}_m\widetilde{\mathbf H}_m^{\top}=mI_m,
\qquad
\mathbf H_m=\frac{1}{\sqrt m}\widetilde{\mathbf H}_m,
\\
&\mathbf H_m\mathbf H_m^{\top}
=\mathbf H_m^{\top}\mathbf H_m=I_m.
\label{eq:supp_hadamard_properties}
\end{align}
For power-of-two orders, the Sylvester construction is
\begin{equation}
\widetilde{\mathbf H}_1=[1],
\qquad
\widetilde{\mathbf H}_{2m}
=
\begin{bmatrix}
\widetilde{\mathbf H}_m & \widetilde{\mathbf H}_m\\
\widetilde{\mathbf H}_m & -\widetilde{\mathbf H}_m
\end{bmatrix}.
\label{eq:supp_sylvester}
\end{equation}
Its butterfly structure evaluates the unnormalized transform using
\(O(m\log m)\) additions and subtractions.

To accommodate channel dimensions close to non-power-of-two orders, we also
use Paley constructions~\cite{steepleton2019constructions,seberry2020hadamard}.
For an odd prime power \(q\), Paley-I provides order \(q+1\) when
\(q\equiv3\pmod 4\), whereas Paley-II provides order \(2(q+1)\) when
\(q\equiv1\pmod 4\). Kronecker products of available matrices provide further
valid orders. We denote the resulting set of supported orders by
\(\mathcal M_H\); no existence assumption is made for unsupported orders.

For a channel dimension \(C\), the transform order is selected as
\begin{equation}
C_4(C)=4\left\lceil\frac{C}{4}\right\rceil,
\qquad
m(C)=\min\left\{m\in\mathcal M_H\mid m\geq C_4(C)\right\}.
\label{eq:supp_hadamard_order}
\end{equation}

\subsection{Padding for DH and WH}

Let \(m=m(C)\geq C\) and define
\begin{align}
&\mathbf P_C=
\begin{bmatrix}
I_C & 0
\end{bmatrix}
\in\mathbb R^{C\times m},
\nonumber\\
&\mathbf T_C^{\mathrm{in}}=\mathbf P_C\mathbf H_m,
\qquad
\mathbf T_C^{\mathrm{out}}=\mathbf P_C\mathbf H_m^{\top}.
\label{eq:supp_padded_transforms}
\end{align}
Both transforms have orthonormal rows:
\begin{equation}
\mathbf T_C^{\mathrm{in}}
\left(\mathbf T_C^{\mathrm{in}}\right)^{\top}
=
\mathbf T_C^{\mathrm{out}}
\left(\mathbf T_C^{\mathrm{out}}\right)^{\top}
=I_C.
\label{eq:supp_semi_orthogonality}
\end{equation}
The orientations in Eq.~\eqref{eq:supp_padded_transforms} agree with the
input- and output-side Hadamard transforms used for DH and WH, respectively.

For a fully connected layer, set
\(\mathbf T_l^{\mathrm{in}}
=\mathbf T_{C_{\mathrm{in},l}}^{\mathrm{in}}\).
For a \(K_h\times K_w\) convolution after patch unfolding, set
\begin{equation}
\mathbf T_l^{\mathrm{in}}
=
I_{K_hK_w}\otimes
\mathbf T_{C_{\mathrm{in},l}}^{\mathrm{in}}.
\label{eq:supp_convolution_input_transform}
\end{equation}
In both cases,
\(\mathbf T_l^{\mathrm{in}}
(\mathbf T_l^{\mathrm{in}})^{\top}=I_{D_l}\).
The padded DH identity is therefore
\begin{equation}
Y_l
=
\left(X_l\mathbf T_l^{\mathrm{in}}\right)
\left(W_l\mathbf T_l^{\mathrm{in}}\right)^{\top}
+b_l.
\label{eq:supp_padded_dh}
\end{equation}
Thus, padding expands only the contracted representation and does not change
the full-precision operator.

For WH, let
\(\mathbf T_l^{\mathrm{out}}
=\mathbf T_{C_{\mathrm{out},l}}^{\mathrm{out}}\) and define
\begin{equation}
\overline W_l
=
\left(\mathbf T_l^{\mathrm{out}}\right)^{\top}W_l,
\qquad
\overline b_l
=
b_l\mathbf T_l^{\mathrm{out}}.
\label{eq:supp_padded_wh_parameters}
\end{equation}
Using Eq.~\eqref{eq:supp_semi_orthogonality}, the WH identity is
\begin{equation}
Y_l
=
\left(
X_l\overline W_l^{\top}+\overline b_l
\right)
\left(\mathbf T_l^{\mathrm{out}}\right)^{\top}.
\label{eq:supp_padded_wh}
\end{equation}
The intermediate WH output has \(m(C_{\mathrm{out},l})\) channels, and the
final semi-orthogonal transform restores the original
\(C_{\mathrm{out},l}\)-channel output.

\subsection{Deployment Realization}

The transformed weights and, for WH, the transformed bias in
Eqs.~\eqref{eq:supp_padded_dh}--\eqref{eq:supp_padded_wh} are
input-independent and can be stored after offline transformation. An
activation-side DH transform is a fixed channel transform. For convolutional
features, it can be represented by a bias-free \(1\times1\) convolution that
simultaneously performs padding and channel mixing, thereby using standard
graph operators for every order in \(\mathcal M_H\).

For a Sylvester order, a dedicated butterfly implementation can replace the
dense transform. Its normalization \(1/\sqrt m\) is absorbed into the adjacent
quantization scale, leaving additions and subtractions in the transform itself.
Paley-based orders may use the portable fixed-transform representation unless a
specialized kernel is available.

\section{Bias and General Integer Nonlinearities}
\label{sec:supp_integer}

Let \(s_{x,l}\) be the scale of the integer activation operand consumed by
layer \(l\), and let \(s_{w,l,c}\) be the weight scale of accumulator channel
\(c\). Integer multiply--accumulate produces \(a_{\mathrm{int},l,c}\) with
scale
\begin{equation}
s_{A,l,c}=s_{x,l}s_{w,l,c}.
\label{eq:supp_accumulator_scale}
\end{equation}
The following derivation assumes symmetric integer codes. The affine case is
addressed at the end of this section.

\subsection{Integer Basis Conversions}

Let \(m_l^{\mathrm{out}}\) and \(m_{l+1}^{\mathrm{in}}\) denote the actual
Hadamard orders, including padding. The unnormalized output-side operator
\(\mathcal H_l^{\mathrm{out}}\) restores the original output basis after a WH
layer and is the identity after a DH layer. Similarly,
\(\mathcal H_{l+1}^{\mathrm{in}}\) generates the Hadamard-domain input for a DH
next layer and is the identity for a WH next layer. Their normalization factors
are
\begin{align}
\gamma_l^{\mathrm{out}}
&=
\begin{cases}
1, & l\text{ is DH},\\
\sqrt{m_l^{\mathrm{out}}}, & l\text{ is WH},
\end{cases}
\nonumber\\
\gamma_{l+1}^{\mathrm{in}}
&=
\begin{cases}
\sqrt{m_{l+1}^{\mathrm{in}}}, & l+1\text{ is DH},\\
1, & l+1\text{ is WH}.
\end{cases}
\label{eq:supp_normalization_factors}
\end{align}
The unnormalized transforms use only additions and subtractions. When padding
is present, they also apply the projections implicit in
Eq.~\eqref{eq:supp_padded_transforms}.

\subsection{Bias Handling}

For DH, define \(\overline b_l=b_l\). For WH, the accumulator is in the
output Hadamard basis, so \(\overline b_l\) is the transformed bias in
Eq.~\eqref{eq:supp_padded_wh_parameters}. The bias code for accumulator
channel \(c\) is
\begin{equation}
b_{\mathrm{int},l,c}
=
\operatorname{round}
\left(
\frac{\overline b_{l,c}}{s_{A,l,c}}
\right).
\label{eq:supp_bias_quantization}
\end{equation}
It is added to the integer accumulator before output-basis conversion and
requantization.

\subsection{General Nonlinearities}

Let \(\phi\) be a nonhomogeneous activation, such as SiLU, GELU, GDN, or IGDN.
Because a common scale factor cannot be propagated through \(\phi\), we
introduce an integer preactivation \(z_{\mathrm{int},l}\) with scale
\(s_{z,l}\) and an integer activation output \(u_{\mathrm{int},l}\) with scale
\(s_{u,l}\).

Choose integer multipliers and shifts such that
\begin{equation}
\frac{s_{A,l,c}}
{\gamma_l^{\mathrm{out}}s_{z,l}}
\approx
\frac{M_{l,c}^{\mathrm{out}}}
{2^{n_l^{\mathrm{out}}}}.
\label{eq:supp_general_output_scale}
\end{equation}
Because accumulator scales can differ across channels,
\(\boldsymbol M_l^{\mathrm{out}}\) is applied before the output-side transform:
\begin{equation}
z_{\mathrm{int},l}
=
\operatorname{clamp}
\left(
\operatorname{round}
\left[
\frac{
\mathcal H_l^{\mathrm{out}}
\left(
\boldsymbol M_l^{\mathrm{out}}\odot a_{\mathrm{int},l}
\right)}
{2^{n_l^{\mathrm{out}}}}
\right],
q_{\min},q_{\max}
\right).
\label{eq:supp_general_preactivation}
\end{equation}
Applying the transform before channel-wise scale alignment would mix integer
values representing different real-valued scales.

The integer nonlinear operator implements
\begin{align}
u_{\mathrm{int},l}
&=
\phi_{\mathrm{int}}
\left(
z_{\mathrm{int},l};s_{z,l},s_{u,l}
\right)
\nonumber\\
&\approx
\operatorname{clamp}
\left(
\operatorname{round}
\left[
\frac{\phi\left(s_{z,l}z_{\mathrm{int},l}\right)}
{s_{u,l}}
\right],
q_{\min},q_{\max}
\right).
\label{eq:supp_general_activation}
\end{align}
An integer lookup table, piecewise approximation, or fixed-point kernel can
realize \(\phi_{\mathrm{int}}\) without materializing an intermediate
real-valued tensor.

The activation is then converted to the operand scale \(s_{x,l+1}\) of the
next layer. Choose
\begin{equation}
\frac{s_{u,l}}
{\gamma_{l+1}^{\mathrm{in}}s_{x,l+1}}
\approx
\frac{M_l^{\mathrm{in}}}
{2^{n_l^{\mathrm{in}}}},
\label{eq:supp_general_input_scale}
\end{equation}
which gives
\begin{equation}
x_{\mathrm{int},l+1}
=
\operatorname{clamp}
\left(
\operatorname{round}
\left[
\frac{
\mathcal H_{l+1}^{\mathrm{in}}
\left(
M_l^{\mathrm{in}}u_{\mathrm{int},l}
\right)}
{2^{n_l^{\mathrm{in}}}}
\right],
q_{\min},q_{\max}
\right).
\label{eq:supp_general_input_transition}
\end{equation}
For a WH next layer,
\(\mathcal H_{l+1}^{\mathrm{in}}=\mathcal I\) and
\(\gamma_{l+1}^{\mathrm{in}}=1\). For a DH next layer, the integer input-side
Hadamard transform and its actual order are used.

Under affine activation quantization, each multiply--accumulate and Hadamard
transform uses the centered code \(x_{\mathrm{int}}-z_x\), and the destination
zero-point is added after requantization. The Hadamard normalization factors
remain absorbed into the fixed-point multiplier--shift pairs. Intermediate
accumulator widths must cover the growth of the unnormalized transforms to
avoid overflow.

\vfill

\end{document}